\documentclass[aps,pre,twocolumn,superscriptaddress,showpacs,floatfix]{revtex4-1}
\usepackage{graphicx}
\usepackage{amssymb,amsmath,amsbsy,amsfonts,bm}
\usepackage{epsfig}
\usepackage{multirow}
\usepackage{threeparttable}
\usepackage[percent]{overpic}
\usepackage{color}

\begin{document}
\title{Flow heterogeneities in supercooled liquids and glasses under shear}
\author{Mehrdad Golkia}
\affiliation{Institut f\"ur Theoretische Physik II: Weiche Materie, Heinrich
Heine-Universit\"at D\"usseldorf, Universit\"atsstra{\ss}e 1, 
40225 D\"usseldorf, Germany}
\author{Gaurav P. Shrivastav}
\affiliation{Institute for Theoretical Physics, TU Wien, Wiedner 
Hauptstra{\ss}e 8-10, 1040 Wien, Austria}
\author{Pinaki Chaudhuri}
\affiliation{The Institute of Mathematical Sciences, IV Cross Road, CIT 
Campus, Taramani, Chennai 600 113, Tamil Nadu, India}
\author{J\"urgen Horbach}
\affiliation{Institut f\"ur Theoretische Physik II: Weiche Materie, Heinrich
Heine-Universit\"at D\"usseldorf, Universit\"atsstra{\ss}e 1, 
40225 D\"usseldorf, Germany}

\begin{abstract}
Using extensive non-equilibrium molecular dynamics simulations, we
investigate a glassforming binary Lennard-Jones mixture under shear.
Both supercooled liquids and glasses are considered. Our focus is
on the characterization of inhomogeneous flow patterns such as shear
bands that appear as a transient response to the external shear.
For the supercooled liquids, we analyze the crossover from Newtonian
to non-Newtonian behavior with increasing shear rate $\dot{\gamma}$.
Above a critical shear rate $\dot{\gamma}_c$ where a non-Newtonian
response sets in, the transient dynamics are associated with the
occurrence of short-lived vertical shear bands, i.e.~bands of high
mobility that form perpendicular to the flow direction.  In the
glass states, long-lived horizontal shear bands, i.e.~bands of high
mobility parallel to the flow direction, are observed in addition
to vertical ones.  The systems with shear bands are characterized
in terms of mobility maps, stress-strain relations, mean-squared
displacements, and (local) potential energies.  The initial formation
of a horizontal shear band provides an efficient stress release,
corresponds to a local minimum of the potential energy, and is
followed by a slow broadening of the band towards the homogeneously
flowing fluid in the steady state. Whether a horizontal or a vertical
shear band forms cannot be predicted from the initial undeformed
sample.  Furthermore, we show that with increasing system size the
probability for the occurrence of horizontal shear bands increases.
\end{abstract}

\maketitle

\section{Introduction}
The structural relaxation in glassforming liquids is associated
with a rapidly growing time scale $\tau$. At the glass transition,
$\tau$ exceeds the experimentally accessible time scale and the
liquid becomes an amorphous solid with an essentially frozen liquid
structure \cite{glassbook}. The slow relaxation processes in the
vicinity of the glass transition make the liquid very sensitive to
even very small external fields. If a highly viscous liquid is,
e.g., sheared with a constant shear rate $\dot{\gamma}$, a new time
scale is introduced by $1/\dot{\gamma}$. If this time scale becomes
smaller than the time scale for structural relaxation, i.e.~$\dot{\gamma}
\tau > 1$, there is a qualitative change of the response of the
liquid to the shear in that it becomes a non-Newtonian liquid
\cite{chhabra}, with the steady-state shear stress, $\sigma_{\rm
ss}$, being no longer proportional to the shear rate $\dot{\gamma}$.
Structural relaxation processes change qualitatively in the
non-Newtonian regime, which is reflected in a decrease of the
effective shear viscosity (shear thinning) as well as anisotropies
in the structure and dynamics of the sheared liquid
\cite{berthier2002,fuchs2005,varnik2006,zausch2008,zausch2009}.

The transient response of a sheared supercooled liquid, i.e.~its
behavior before the steady state is reached, also changes qualitatively
in the non-Newtonian regime. Here, the stress-strain relation shows
a maximum at a strain of the order of 0.1, followed by a stress
relaxation towards $\sigma_{\rm ss}$ \cite{zausch2008}.  A similar
behavior is also seen for a glass system in response to an external
shear at a constant shear rate. At any finite shear rate, the glass
is expected to undergo a transformation from a deformed amorphous
solid to a homogeneously flowing fluid in the steady state (provided
one waits long enough until the steady state is reached). However,
below the glass transition, a Newtonian regime can no longer be
seen on experimentally accessible time scales and the extrapolation
towards vanishing shear rates, $\dot{\gamma} \to 0$, suggests the
approach of a constant value of $\sigma_{\rm ss}$ which can be
interpreted as a yield stress.  Thus, in this context, the occurrence
of a yield stress is a kinetic phenomenon.  It is associated with
the response of a non-equilibrium glass state and the accessible
time scales in experiments (or simulations).  One should keep in
mind, however, that the ``natural response'' of an amorphous system
to shear is that of a Newtonian fluid in the limit $\dot{\gamma}\to 0$.

The transient behavior of sheared glass can be markedly different
from that of a supercooled liquid with respect to 
the spatial response to the shear field. In supercooled
liquids, there are inhomogeneous flow patterns prior to the steady
state, albeit these features are usually very weak \cite{fuereder2017}.
In glasses, however, such inhomogeneities are very pronounced. In
particular, there is the possibility for the formation of shear
bands, i.e.~band-like structures with a higher strain or mobility
than in other regions of the system. The occurrence of shear bands
is of technological relevance since they can cause the mechanical
failure of a glassy material. Shear bands have been observed in
experiments of soft matter systems \cite{besseling2010, divoux2010,
chikkadi2011, divoux2016} and metallic glasses \cite{schuh2007,
maass2015, bokeloh2011, binkowski2016} as well as in computer
simulations of various model glass systems \cite{varnik2003,
bailey2006, shi2006, shi2007, ritter2011, sopu2011, chaudhuri2012,
dasgupta2012, dasgupta2013, albe2013, shiva2016_2, ozawa2018, ozawa2019,
singh2019, parmar2019, miyazaki2019}.

Whether or not shear bands form, depends on the degree of annealing
of the initial undeformed glass sample and the applied shear rate.
This has been recently demonstrated by Ozawa {\it et al.}~\cite{ozawa2018,
ozawa2019} using an athermal quasistatic shear (aqs) protocol.
These authors have also shown that shear bands with an orientation
parallel to the flow direction (in the following denoted as horizontal
shear bands) are initiated by a very pronounced stress drop in the
stress-strain relation. Moreover, in analogy to a first-order phase
transition, the stress drop becomes sharper with increasing system
size. This has led to the interpretation that the stress drop in
the stress-strain relation indicates the presence of the spinodal
of a first-order non-equilibrium phase transition \cite{ozawa2018,
procaccia2017}. However, in this picture, the coexisting
phases underlying the proposed first-order transition are not identified.
As a matter of fact, the inhomogeneous state with shear band
is unstable. With further deformation the shear band grows until
the system is fully fluidized (see also below). Thus, there is a
``transition'' from a deformed glass to a homogeneously flowing
fluid and the shear-banded structures appear to be transient.

At this point it is interesting to compare the response of glasses
and crystals to a deformation. It is tempting to see an analogy of
the shear bands in glasses to slip planes in crystals. Slip planes
in crystals are located between layers of particles and thereby do
not contain any particles. They allow to fully release the stresses
in response to a deformation (at least if one considers crystalline
systems in the thermodynamic limit) \cite{sausset2010, nath2018,
reddy2020}.  Unlike a crystal, an amorphous solid can never fully
release the stresses in response to a deformation. The stresses
that might be localized in a shear band are always carried by
particles and, as mentioned above, inhomogeneous systems with shear
bands are not stable and transform into a homogeneously flowing
fluid state with further deformation.  The scenario is completely
different in a crystal. Here, at sufficiently small deformation
rates, the flow of the crystal can be characterized as a repeated
visit of stress-free states \cite{nath2018, vanmegen1976}.

However, contrary to the first-order transition scenario proposed
by Ozawa {\it et al.}~which requires the existence of overhangs in
the stress-strain relation, Barlow {\it et al.}~\cite{barlow} have
demonstrated using mesoscale constitutive models, that, with
increasing age, there can be a gradual change from ductile to
brittle-like behaviour. For increased annealing, it has been shown
in the framework of mesoscale models \cite{moorcroft2013} that there
is an increase of the height of the stress overshoot, associated
with an increasingly sharper decay towards the steady state stress.
In this scenario, when the slope of the latter decay becomes largely
negative for more annealed states, a mechanical instability kicks
in and shear banding ensues, with the process becoming more
catastrophic with larger and larger slopes for more and more aging.
In contrast, for the younger ductile materials, where the stress
decay is softer, no such flow heterogeneities occur.

In this work, we are aiming at a characterization of inhomogeneous
flow patterns in glasses.  We use non-equilibrium molecular dynamics
(MD) simulations to investigate planar Couette flow in a binary
Lennard-Jones system. As a reference, we first consider supercooled
liquids for which we analyze the crossover from Newtonian to
non-Newtonian behavior with increasing shear rate $\dot{\gamma}$.
In the non-Newtonian regime, the stress-strain relations show a
stress drop from a maximum stress to the steady-state stress,
$\sigma_{\rm ss}$, that decreases with decreasing shear rate and
vanishes logarithmically at a critical shear rate $\dot{\gamma}_c$.
For $\dot{\gamma}<\dot{\gamma}_c$, the response of a Newtonian fluid
with $\sigma_{\rm ss}\propto \dot{\gamma}$ is seen. After the stress
drop, we find the occurrence of vertical shear bands in the transient
response of the non-Newtonian fluid. These flow patterns, that form
in the direction perpendicular to the flow, are short-lived and can
be seen as signatures of the inhomogeneities, observed in the
transient response of glasses.

For our study, we consider very low temperature
glassy states, where although thermal fluctuations are present,  
the affine displacements due to the applied deformation
dominate over the thermally induced random motion of the particles. This is reflected
in a ballistic regime $\propto \dot{\gamma}^2 t^2$ (with $t$ the
time) in the mean-squared displacements before the plastic flow sets in.
In the considered glasses under shear, we observe in addition to
samples with short-lived vertical shear bands, samples with long-lived
horizontal shear bands. After their formation, the latter bands
exhibit a slow broadening with increasing strain until the system
reaches the steady state. Furthermore, it is not
imprinted in the initial undeformed glass sample, whether and where horizontal
bands form. Instead, we find that their formation is linked to
stochasticity; small changes in the protocol such as the use of
different initial random numbers for the thermostat can change the
flow patterns from vertical to horizontal bands and vice versa.
When a system-spanning horizontal shear band is first nucleated
after a relatively sharp stress drop, the total potential energy
of the system exhibits a minimum with a value which is below that
reached in the steady state. So in the case of horizontal shear
bands, the transition of the deformed amorphous solid to the
homogeneously flowing fluid state takes place via an unstable
intermediate state that has, however, a lower potential energy than
the final steady state.

The rest of the paper is organized as follows: In the next section,
we present the details of the model potential and the MD simulations.
Then, in Sec.~\ref{sec_results}, the results for the supercooled
liquids and glasses under shear are presented. Finally, we draw
conclusions in Sec.~\ref{sec_conclusions}.

\section{Model and simulation details}
\label{sd}
We consider the Kob-Andersen (KA) binary Lennard-Jones mixture
\cite{kob1994} that has been widely used in many computer simulation
studies as a model for a glassforming system. The mixture consists
of 80\% A particles and 20\% B particles. The interaction potential
between a particle of type $\alpha$ and a particle of type $\beta$
($\alpha, \beta = \textrm{A, B}$), separated by a distance $r\le
R_c$, is given by
\begin{eqnarray}
\label{LJ1}
u_{\alpha\beta}(r) &=& 
\phi_{\alpha\beta}(r)-\phi_{\alpha\beta}(R_{c})-\left(r-R_{c}\right)\left. 
\frac{d\phi_{\alpha\beta}}{dr}\right|_{r=R_{c}},\nonumber\\
\phi_{\alpha\beta}(r) &=& 
4\epsilon_{\alpha\beta}\left[\left(\sigma_{\alpha\beta}/r\right)^{12}-
\left(\sigma_{\alpha\beta}/r\right)^{6}\right] \, .
\end{eqnarray}
The values of the interaction parameters are set to $\epsilon_{\textrm{AA}}
= 1.0$, $\epsilon_{\textrm{AB}} = 1.5\epsilon_{\textrm{AA}}$,
$\epsilon_{\textrm{BB}} = 0.5\epsilon_{\textrm{AA}}$, $\sigma_{\textrm{AA}}
= 1.0$, $\sigma_{\textrm{AB}} = 0.8\sigma_{\textrm{AA}}$, and
$\sigma_{\textrm{BB}} = 0.88\sigma_{\textrm{AA}}$. In the following,
we use $\epsilon_{\textrm{AA}}$ and $\sigma_{\textrm{AA}}$ as the
unit for energy and length, respectively.  The cutoff radius in
Eq.~(\ref{LJ1}) is chosen as $R_{c} = 2.5\sigma_{\textrm{AA}}$.  As
the time unit, we use $\sqrt{{m\sigma_{\textrm{AA}}^{2}} /
\epsilon_{\textrm{AA}}}$, where $m$ is the mass of a particle that
is considered to be equal for both type of particles, i.e.~$m =
m_{\textrm{A}} = m_{\textrm{B}} = 1.0$.  More details about the
model can be found in Ref.~\cite{kob1994}.

We perform the non-equilibrium MD simulation at constant particle
number $N$, constant volume $V$, and constant temperature $T$, using
the LAMMPS package \cite{plimpton1995}.  The systems, be they under
shear or without shear, are thermostatted via dissipative particle
dynamics (DPD) \cite{soddemann2003}.  The DPD equations of motions
are as follows:
\begin{eqnarray}
\label{neq}
\dot{\bm r}_{i} &=& {\bm p}_{i}/m_{i},\\
\label{dpdeq}
\dot{\bm p}_{i} &=& \sum_{j\neq i} \left[{\bm F}_{ij} 
+ {\bm F}^{\rm D}_{ij} + {\bm F}^{\rm R}_{ij}\right], 
\end{eqnarray}
with ${\bm r}_{i}$ the position and ${\bm p}_{i}$ the momentum of
a particle $i$. Further, ${\bm F}_{ij}$ represents the conservative
force on a particle pair $i, j$ due to the interparticle interaction,
defined by Eq.~(\ref{LJ1}). The dissipative force, ${\bm F}^{\rm
D}_{ij}$, is given by
\begin{eqnarray}
\label{dissf}
{\bm F}^{\rm D}_{ij} = -\zeta \omega^{2}\left(r_{ij}\right) 
\left(\hat{\bm r}_{ij}\cdot {\bm v}_{ij}\right)\hat{\bm r}_{ij}, 
\end{eqnarray}
with $\zeta$ the friction coefficient, ${\bm r}_{ij}$ the distance
vector between particles $i$ and $j$, $\hat{\bm r}_{ij}$ the unit
vector of ${\bm r}_{ij}$ , $r_{ij}$ the distance between the two
particles, and ${\bm v}_{ij} = {\bm v}_{i} - {\bm v}_{j}$ the
relative velocity between them.

The value of $\zeta$ is chosen to be equal to $1$. Furthermore, for
$\omega\left(r_{ij}\right)$, the following function is used:
\begin{eqnarray}
\label{omega}
\omega (r_{ij}) = 
\begin{cases}
\sqrt{1-r_{ij}/r_{c}} & \quad if~ r_{ij}< r_{c},\\
0                     & \quad {\rm otherwise}.
\end{cases}
\end{eqnarray}
The cutoff radius for this function, $r_{c}$, is taken equal to the
cutoff radius of the interaction potential, $R_{c}$. In Eq.~(\ref{dpdeq}),
$F^{\rm R}_{ij}$ represents the random force which is defined as
\begin{eqnarray}
\label{ranf}
F^{\rm R}_{ij} = 
\sqrt{2k_{B}T\zeta}\omega (r_{ij})\theta_{ij}\hat{\bm r}_{ij}\, .
\end{eqnarray}
Here, $\theta_{ij}$ are uniformly-distributed random numbers with
zero mean and unit variance. For further details about the DPD
thermostat parameters, see Refs.~\cite{zausch2008, zausch2009}.
The equation of motion, Eq.~(\ref{neq})-(\ref{dpdeq}), are integrated
via the velocity Verlet algorithm using an integration time step
$\Delta t = 0.005$.

We consider the KA mixture at the two different densities $\rho=1.2$
and $\rho=1.3$ in the supercooled liquid and glass states. For these
densities, the mode coupling glass transition temperatures are at
$T_c=0.435$ and $T_c=0.68$, respectively.  Note that most recent
studies have considered the KA mixture at $\rho=1.2$. However, at
this density, one encounters a glass-gas miscibility gap at low
temperatures \cite{chaudhuri2016_1, chaudhuri2016_2}, which is not
the case for $\rho=1.3$. Therefore, below we only consider supercooled
liquid states at $\rho=1.2$ while at $\rho=1.3$ both supercooled
liquids and glasses at very low temperatures are simulated.  To
prepare the glass samples at $\rho = 1.3$, we start with equilibrated
samples at $T = 0.7$ (in the supercooled regime) and quench it
instantaneously to a glass state at the temperature $T = 10^{-4}$,
followed by aging of the samples over the time $t_{w} = 10^{4}$,
which would correspond to moderate annealing. For the supercooled
liquids at $\rho=1.2$, we have considered cubic samples with linear
size $L = 30$ and 50. At $\rho = 1.3$, cubic samples with linear
dimension $L = 60$ are chosen. To study finite-size effects, we
also report simulations for other system sizes. The details of these
simulations can be found below.

We impose a planar Couette flow on the different bulk supercooled
liquids and glasses by shearing them along the $xz$-plane in the
direction of $x$.  The shear is applied via the boundaries using
Lees-Edwards boundary conditions \cite{lees1972}. The shear rates,
$\dot{\gamma}$, considered in this work range from $\dot{\gamma}=10^{-3}$
to $\dot{\gamma}=10^{-7}$ for the supercooled liquid.  For the glass
state, the shear rate $\dot{\gamma}=10^{-4}$ is chosen.  More details
about the model and simulation protocol can be found in 
\cite{shiva2016_1, shiva2016_2}.

\section{Results}
\label{sec_results}
In the following, we first analyse the transient dynamics of
supercooled liquids under shear. We analyse the crossover from
Newtonian to non-Newtonian response of the liquid with increasing
shear rate. The characterization of non-Newtonian liquids then helps
in the understanding of glass states at very low temperature that
are considered in the second part of this section. Here, the major
issue is the study of transient inhomogeneous flow patterns that
form under the application of the shear before the steady state is
reached. In particular, different types of shear bands perpendicular
and parallel to the flow direction are observed.  These flow patterns
are analysed in terms of (local) potential energies, shear stresses,
and (local) mean-squared displacements (MSDs).

\subsection{Supercooled liquids under shear}
In a Newtonian fluid, the steady-state shear stress depends linearly
on the shear rate, $\sigma_{\rm ss}=\eta \dot{\gamma}$ with $\eta$
the shear viscosity. Deviations from this behavior are expected
around a shear rate $\dot{\gamma}_c$ when the time scale, associated
with this shear rate, is of the order of the time scale $\tau$ for
structural relaxation of the unsheared liquid, i.e.~$\dot{\gamma}_c\tau
\approx 1$. As we shall see below, the non-Newtonian response can
be inferred from the occurrence of an overshoot in the stress-strain
relation.

{\bf Shear stress.} To compute the shear stress, we use the virial
equation, given by
\begin{eqnarray}
\label{str}
\left\langle \sigma_{xz}(t)\right\rangle = 
\frac{1}{V}\left\langle\sum_{i}
\left[m_{i}v_{i,x}v_{i,z} + \sum_{i>j} r_{ij,x}F_{ij,z}\right] 
\right\rangle \, ,
\end{eqnarray}
with $V$ the total volume, $m_{i}$ the mass of the $i^{th}$ particle,
$v_{i,x}$ and $v_{i,z}$ respectively the $x$ and $z$ components of
the velocity of the $i^{th}$ particle, $r_{ij,x}$ the $x$ component
of the displacement vector between particles $i$ and $j$, and
$F_{ij,z}$ the $z$ component of the force between particles $i$ and
$j$. Note that the kinetic terms $\propto m_{i}v_{i,x}v_{i,z}$ are
very small and so we have neglected them in our calculation of the
shear stress (cf.~Ref.~\cite{zausch2009}).

%%%%%%%%%%%%%%%%%%%%%%%%%%%%%%%%%%%%%%%%%%%%
\begin{figure}[!htbp]
\begin{center}
\includegraphics[width=6cm]{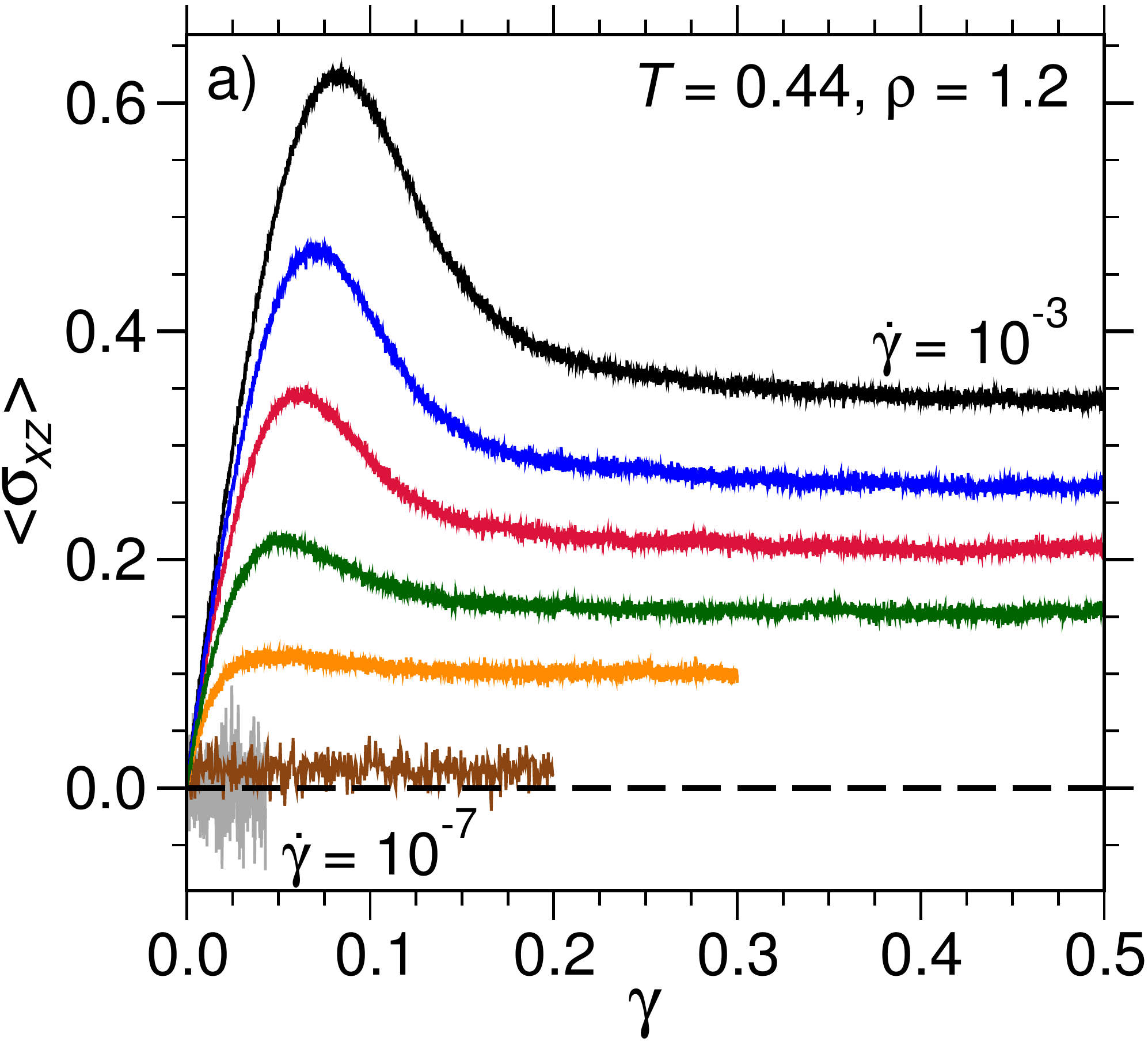}
\includegraphics[width=6cm]{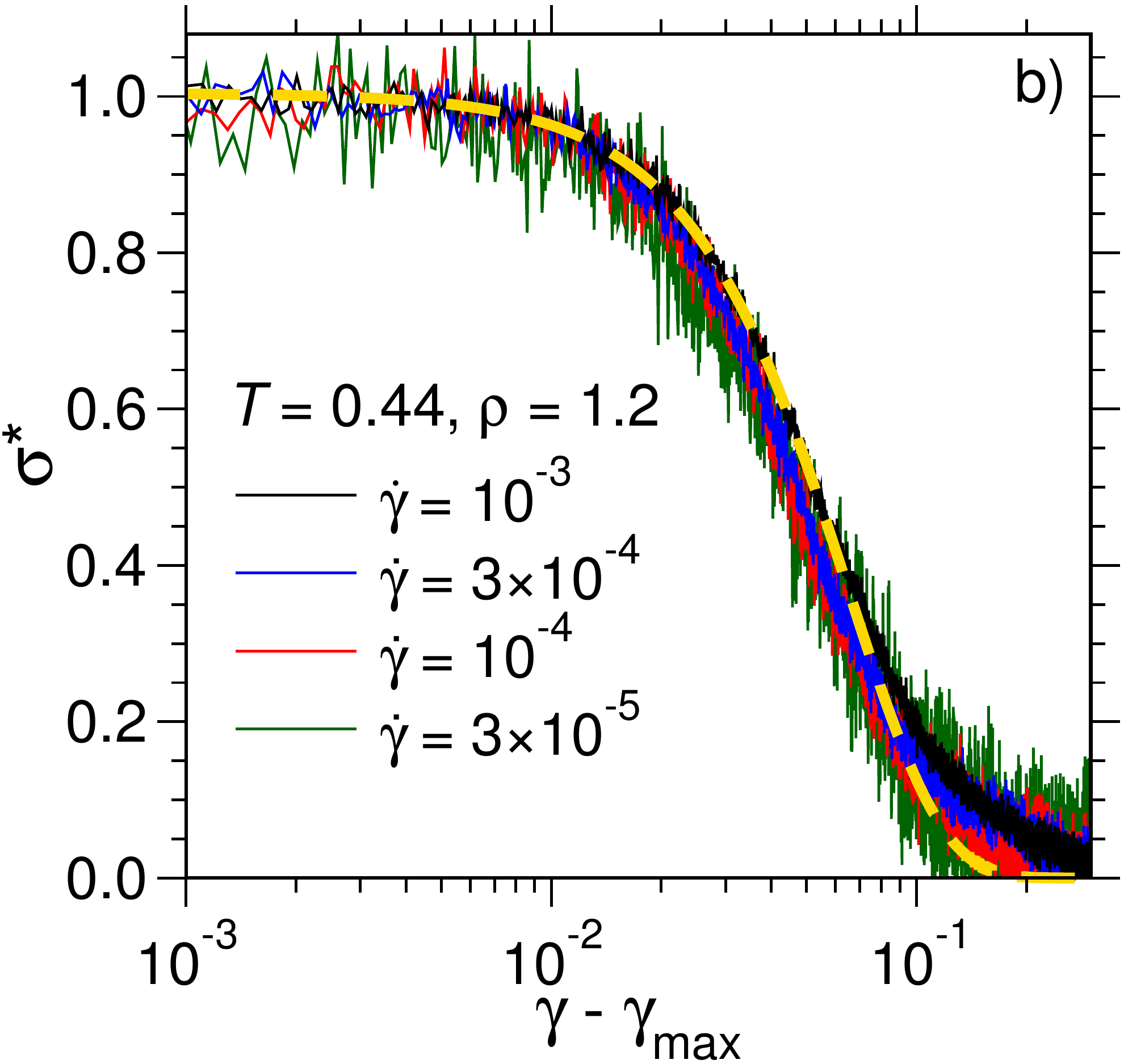}
\caption{\label{fig1} a) Stress-strain relations for the supercooled
liquids at $T = 0.44$ and $\rho = 1.2$ for the shear rates $\dot{\gamma}
= 10^{-3}$ (black), $3\times 10^{-4}$ (blue), $10^{-4}$ (red),
$3\times 10^{-5}$ (green), $10^{-5}$ (orange), $10^{-6}$ (brown),
$10^{-7}$ (grey). As a guide to the eye, a dashed horizontal line
at zero stress is drawn.  b) The reduced stress, $\sigma^\star$,
as obtained by Eq.~(\ref{decay_str}), for different strain rates
in the non-Newtonian regime. The golden dashed line is a fit with
a compressed exponential (see text).}
\end{center}
\end{figure}
%%%%%%%%%%%%%%%%%%%%%%%%%%%%%%%%%%%%%%%%%%%%%%%

In Fig.~\ref{fig1}a, we plot the evolution of the stress, $\langle
\sigma_{xz}\rangle$, for the supercooled liquid at $T = 0.44$ and
$\rho = 1.2$ as a function of strain, $\gamma= \dot{\gamma}t$, for
shear rates $\dot{\gamma} = 10^{-3}$, $3\times 10^{-4}$, $10^{-4}$,
$3\times 10^{-5}$, $10^{-5}$, $10^{-6}$ and $10^{-7}$. At high shear
rates ($\dot{\gamma} = 10^{-3}, 10^{-4}$ and $\dot{\gamma} =
10^{-5}$), the stress increases and reaches a maximum at $\sigma_{\rm
max}$, from where it relaxes to the steady state stress $\sigma_{\rm
ss}$ at large strains. Obviously, the ``stress overshoot'' at
$\sigma_{\rm max}$ decreases with decreasing shear rate and fully
disappears at low shear rates (here, this happens at $\dot{\gamma}
= 10^{-6}$ and $10^{-7}$, see Fig.~\ref{fig1}a).

The emergence of the overshoot in the stress-strain relation marks
the onset of a non-Newtonian response of the liquid.  It is associated
with a relaxation of the stress from $\sigma_{\rm max}$ to $\sigma_{\rm
ss}$.  To analyse this relaxation process, we subtract $\sigma_{\rm
ss}$ from the shear stress $\langle \sigma_{xz}\rangle$ and divide
this difference by $\sigma_{\rm max}$ to obtain the reduced stress
\begin{eqnarray}
\label{decay_str}
\sigma^\star = 
\frac{\langle\sigma_{xz}\rangle - \sigma_{\rm ss}}{\sigma_{\rm max}} \, . 
\end{eqnarray}
In Fig.~\ref{fig1}b, we plot $\sigma^\star$ for different shear
rates in the non-Newtonian regime as a function of $\gamma -
\gamma_{\rm max}$, where $\gamma_{\rm max}$ corresponds to the
strain at the stress $\sigma_{\rm max}$. The curves for the different
shear rates fall roughly onto a master curve that can be fitted
with the compressed exponential function $f\left(x\right) =
\exp\left(-Ax^{B}\right)$, with $x=\gamma-\gamma_{\rm max}$,
$A=15.3822$ and $B=1.69862$ (dashed golden line). Note that the fit
with the compressed exponential does not provide a perfect description
of the data over the whole strain window, but it just gives a rough
idea about functional form of the decay of $\sigma$ from $\sigma_{\rm
max}$ to $\sigma_{\rm ss}$. The decay of the stress in this manner
is intimately related to a superlinear increase of the mean-squared
displacement (see Ref.~\cite{zausch2008} and Fig.~\ref{fig4} below).

Figure \ref{fig1}b also indicates that the strain window over which
the stress is released from the stress maximum towards the steady-state
stress is of the order of $\Delta \gamma = 0.1$. This corresponds
to the typical strain required for the breaking of cages around
each particle in the supercooled regime above the mode coupling
temperature $T_c$ \cite{zausch2008}. Below, we will see that the
decay of $\sigma^\star$ for glass states at very low temperatures
typically occurs on a smaller strain window.

%%%%%%%%%%%%%%%%%%%%%%%%%%%%%%%%%%%%%%%%%%%%
\begin{figure}[!htbp]
\begin{center}
\includegraphics[width=6cm]{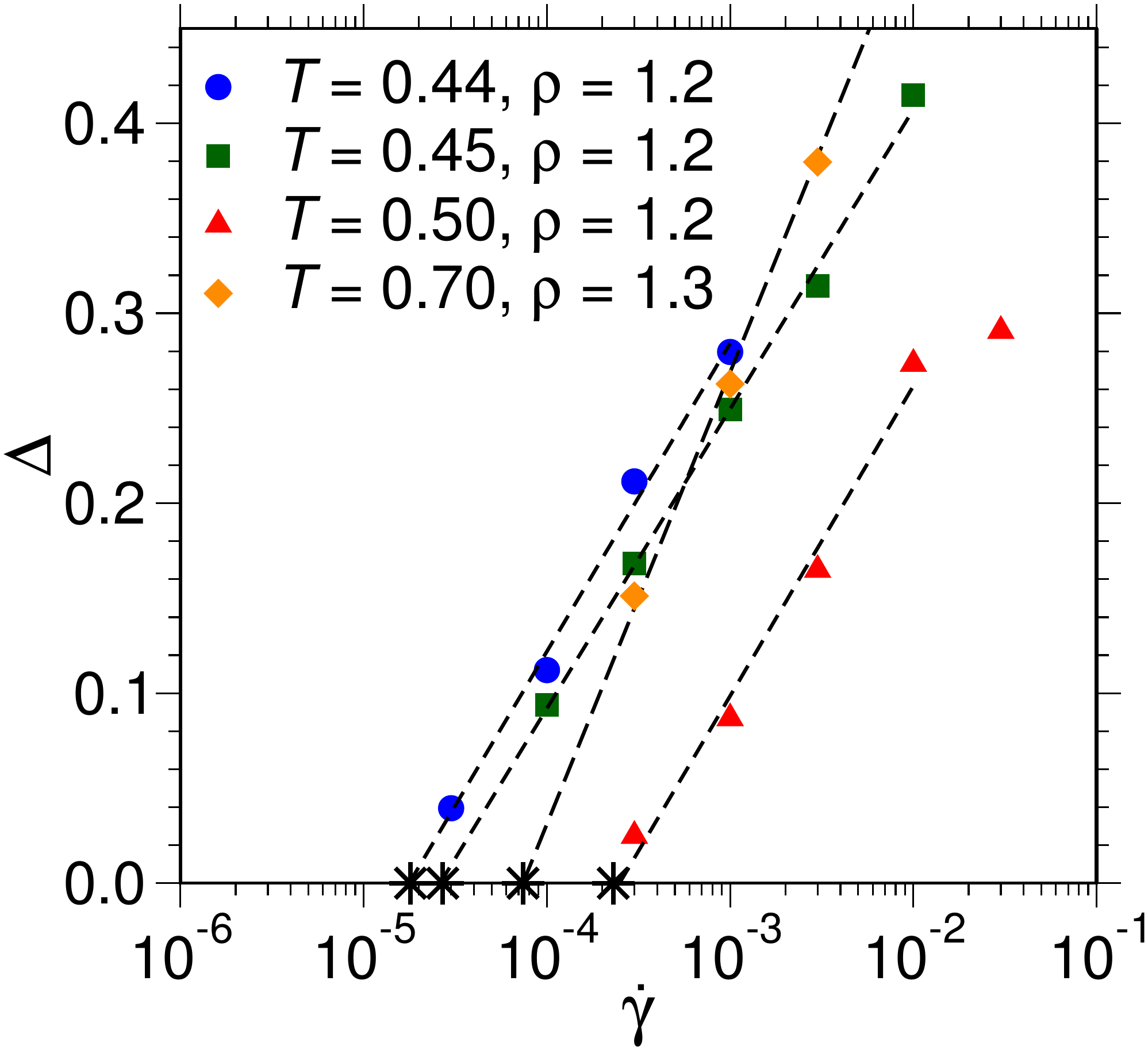}
\caption{\label{fig2} Stress drop $\Delta= \sigma_{\rm max} -
\sigma_{\rm ss}$ as a function of shear rate for the indicated
densities and temperatures.  The dashed lines are fits with the
function $f(\dot{\gamma})=A \, {\rm log} (\dot{\gamma}/\dot{\gamma}_c)$.
The stars mark the values of the critical shear rate $\dot{\gamma}_c$.}
\end{center}
\end{figure}
%%%%%%%%%%%%%%%%%%%%%%%%%%%%%%%%%%%%%%%%%%%%%%%

So we have seen that the non-Newtonian regime is characterized by
the occurrence of a stress drop $\Delta = \sigma_{\rm max} -
\sigma_{\rm ss}$ that decreases with decreasing shear rate and
vanishes below a critical shear rate $\dot{\gamma}_c$. Thus, this
critical shear rate marks the crossover from Newtonian to non-Newtonian
behavior.  In Fig.~\ref{fig2}, the stress drop $\Delta$ is plotted
as a function of ${\rm log} (\dot{\gamma})$ for different temperatures
at $\rho=1.2$ as well as for $T=0.7$ at $\rho=1.3$. This plot
indicates that $\Delta = A \, {\rm log} (\dot{\gamma}/\dot{\gamma}_c)$,
with $A$ and $\dot{\gamma}_c$ being fit parameters, holds at
sufficiently low shear rates.  Here, the critical shear rate
$\dot{\gamma}_c$ corresponds to the value of $\dot{\gamma}$ at
$\Delta = 0$. From the fits, we obtain for $\rho = 1.2$ the values
$\dot{\gamma}_c = 1.8\times10^{-5}$, $2.7\times10^{-5}$, and
$2.3\times10^{-4}$ for $T=0.44$, $T=0.45$, and $T=0.5$, respectively,
and for $\rho = 1.3$ and $T=0.7$ the value $\dot{\gamma}_c =
7.4\times10^{-5}$. In Fig.~\ref{fig2}, these values are marked by
stars.

%%%%%%%%%%%%%%%%%%%%%%%%%%%%%%%%%%%%%%%%%%%%
\begin{figure}%[!htbp]
\begin{center}
\includegraphics[width=6.cm]{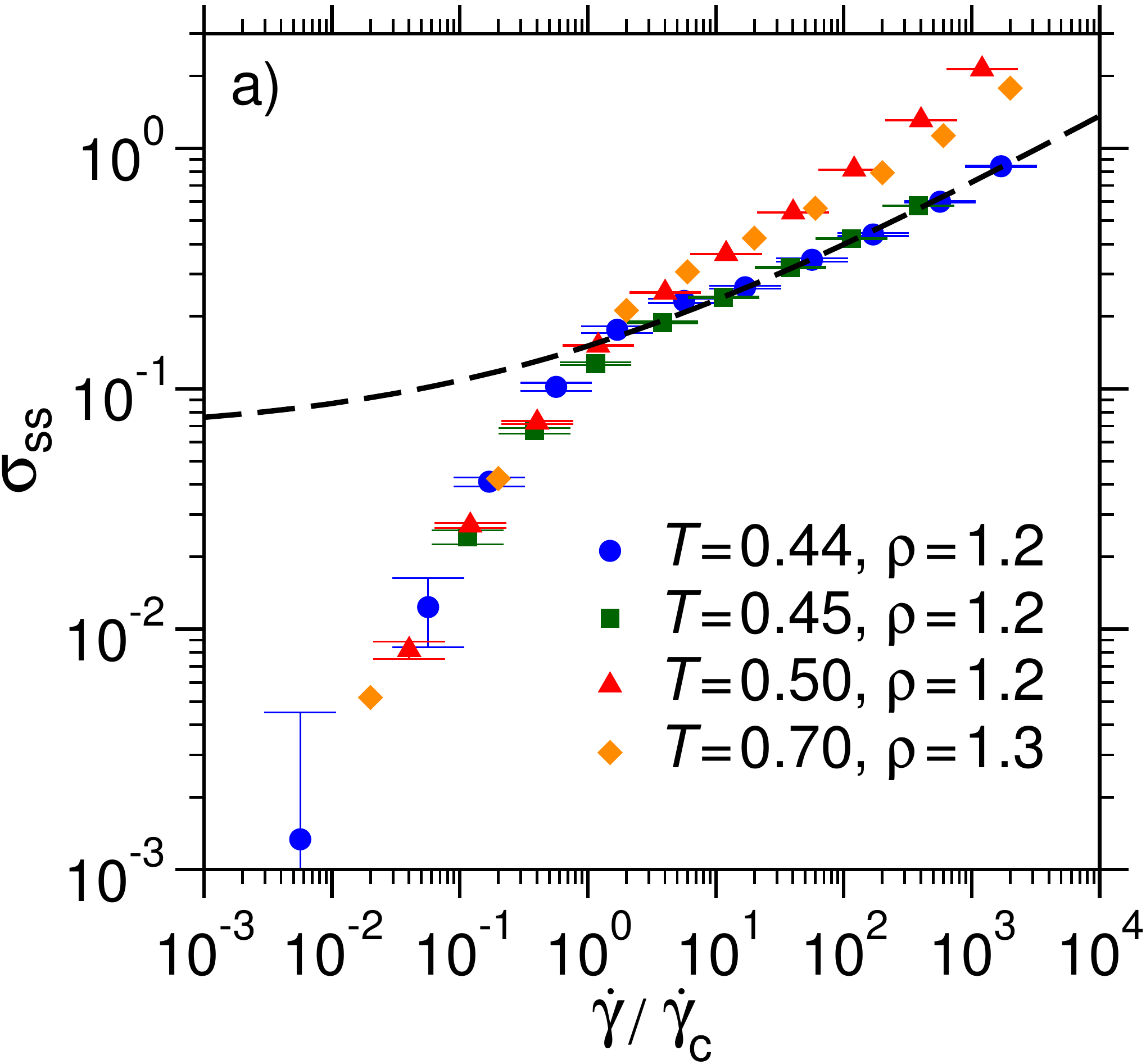}
\includegraphics[width=6.cm]{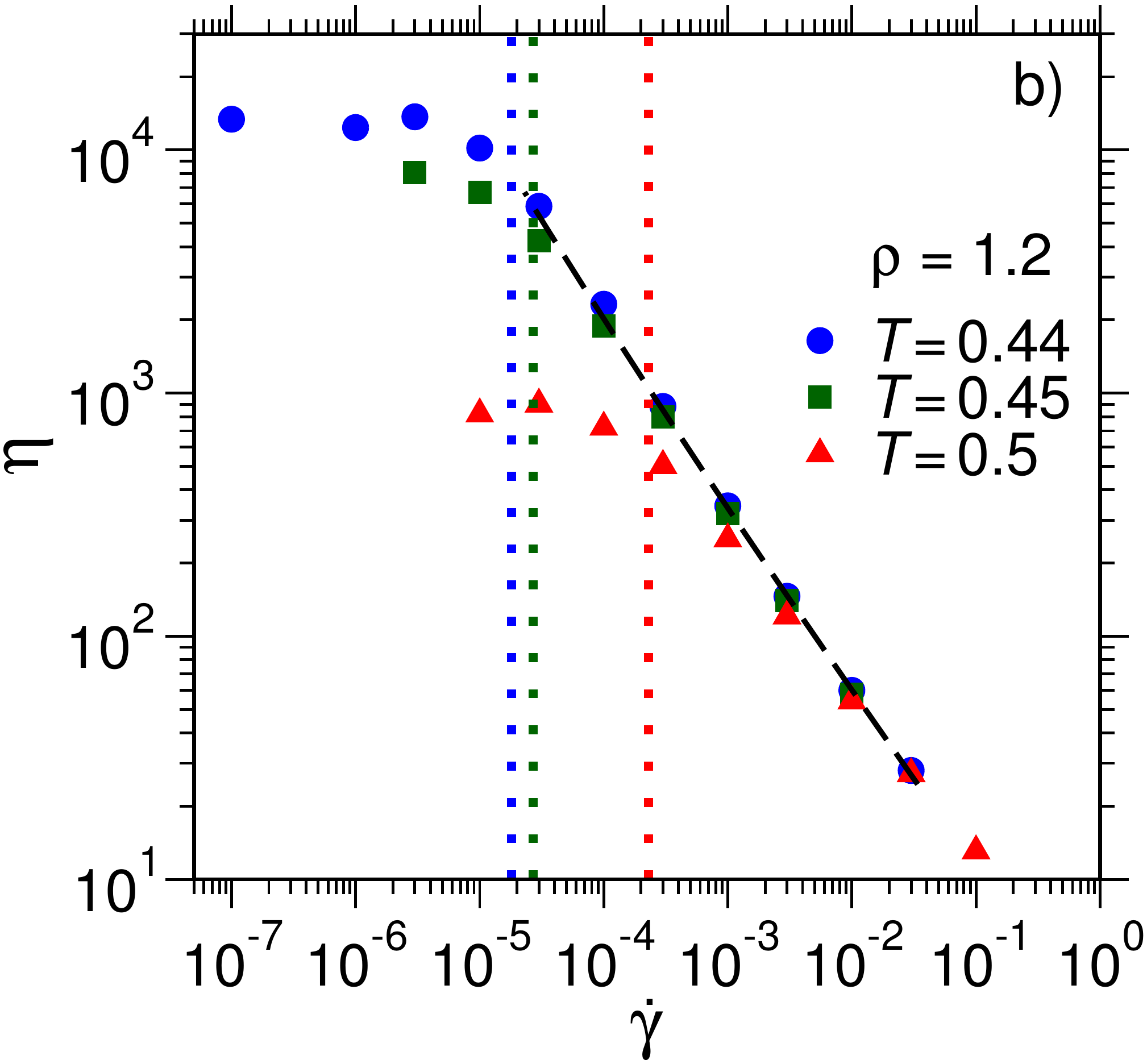}
\caption{\label{fig3} a) The flow curves for the supercooled liquids
for the temperatures $T = 0.44$, 0.45, and 0.5 at $\rho = 1.2$ as
well as for the temperature $T = 0.7$ at $\rho = 1.3$. The dashed
line is a fit with a Herschel-Bulkley law (see text).  b) The
corresponding viscosity $\eta=\sigma_{\rm ss}/\dot{\gamma}$ as a
function of $\dot{\gamma}$ for the data at $\rho = 1.2$. The black
dashed line corresponds to the Herschel-Bulkley law from a), converted
to $\eta(\dot{\gamma})$. The dotted vertical lines with different 
colors mark the values of $\dot{\gamma}_c$ corresponding to the 
different temperatures.}
\end{center}
\end{figure}
%%%%%%%%%%%%%%%%%%%%%%%%%%%%%%%%%%%%%%%%%%%%%%%

Around the critical shear rate $\dot{\gamma}_c$, the flow curve,
i.e.~the shear rate dependence of the steady-state stress $\sigma_{\rm
ss}$, is expected to change from a linear to a non-linear function.
Figure~\ref{fig3}a shows $\sigma_{\rm ss}$ as a function of
$\dot{\gamma}/\dot{\gamma}_c$ for $T=0.44$, 0.45, and 0.5 at
$\rho=1.2$ as well as for $T = 0.7$ at $\rho=1.3$. In the Newtonian
regime, the flow curves $\sigma_{\rm ss}(\dot{\gamma}/\dot{\gamma}_c)$
collapse on top of each other. A non-linear behavior is seen for
$\dot{\gamma}/\dot{\gamma}_c>1$, i.e.~in the non-Newtonian regime.
Moreover, in the latter regime, the different curves do not fall
onto a master curve which indicates that the characteristic time
scale, corresponding to the non-Newtonian regime, is no longer the
time scale $\tau$ for structural relaxation ($\alpha$ relaxation)
of the unsheared liquid. Now, $\dot{\gamma} > 1/\tau$ holds and
thus the time scale $1/\dot\gamma$ matches relaxation time scales
before and during the break-up of the cages around particles that
are formed by the other particles. Note that this time regime is
often called $\beta$ relaxation regime. At very large shear rates,
i.e.~$\dot{\gamma} > 1$, microscopic time scales of the liquid are
probed.

A pronounced $\beta$ relaxation regime (reflected e.g.~by a pleateau
in the mean-squared displacement, see below) is seen in the quiescent
liquid at the temperatures $T=0.44$ and $T=0.45$. For these
temperatures, the flow curves can be fitted in the non-Newtonian
regime to a Herschel-Bulkley law, $\sigma_{\rm ss} = \sigma_{\rm
yield} + A \left(\dot{\gamma}/\dot{\gamma}_c\right)^\alpha$ with
the ``yield stress'' $\sigma_{\rm yield}=0.0648827$, the amplitude
$A=0.0855875$, and the exponent $\alpha=0.295289$ (dashed line in
the main plot of Fig.~\ref{fig3}a) \cite{fn1}. Here, $\sigma_{\rm
yield}$ has to be considered as an effective fit parameter.  However,
for the glass states at temperatures far below $T_c$ the yield
stress $\sigma_{\rm yield}$ indicates the limiting steady-state
stress value, i.e.~the stress will not fall below this value with
respect to the shear rates that are accessible in the simulation.

The transition to the non-Newtonian regime can be also clearly seen
in the behavior of the shear viscosity $\eta = \sigma_{\rm
ss}/\dot{\gamma}$, which is shown in Fig.~\ref{fig3}b as a function
of $\dot{\gamma}$ for the data at $\rho=1.2$. Beyond the Newtonian
regime, where $\eta$ is constant, the shear viscosity decreases with
increasing shear rate, i.e.~we observe shear thinning.  We can
effectively describe the shear thinning behavior for $\dot{\gamma}>
10^{-4}$ by a power law, derived from the Herschel-Bulkley law with
which the flow curves at $T=0.44$ and $T=0.45$ are fitted in
Fig.~\ref{fig3}a.  The corresponding fit is shown as a black dashed
line in Fig.~\ref{fig3}b.

%%%%%%%%%%%%%%%%%%%%%%%%%%%%%%%%%%%%%%%%%%%%
\begin{figure}%[!htbp]
\begin{center}
\includegraphics[width=6.cm]{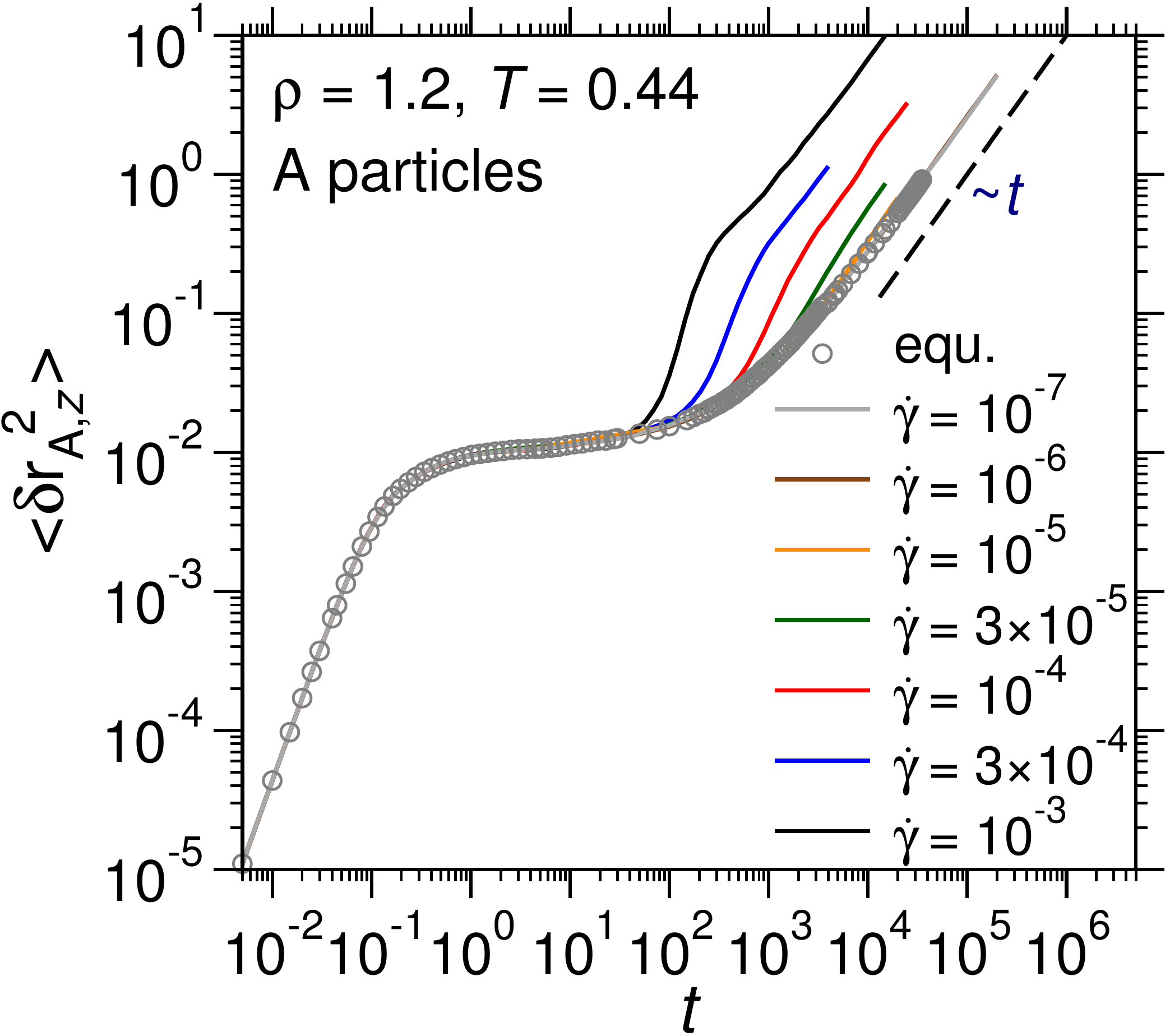}
\caption{\label{fig4} The $z$-component of the MSD of A particles,
$\langle \delta r_{{\rm A}, z}^2 \rangle$, as a function of time
for the supercooled liquid at $\rho = 1.2$ and $T = 0.44$. Data,
corresponding to the equilibrium ($\dot{\gamma}=0$) and the shear
rates $\dot{\gamma} = 10^{-3}$, $3\times 10^{-4}$, $10^{-4}$,
$3\times 10^{-5}$, $10^{-5}$, $10^{-6}$, and $10^{-7}$, are shown.  
The dashed line represents a linear function.}
\end{center}
\end{figure}
%%%%%%%%%%%%%%%%%%%%%%%%%%%%%%%%%%%%%%%%%%%%%%%

{\bf One-particle dynamics.} The crossover from the Newtonian to
the non-Newtonian regime can be also inferred from the one-particle
dynamics. An important quantity that characterizes the dynamics of
a tagged particle of species $\alpha$ ($\alpha = {\rm A, B}$) is
the mean-squared displacement (MSD), defined as
\begin{eqnarray}
\label{msdeq}
\langle\delta r_{\alpha}^{2}(t)\rangle =
\frac{1}{N_{\alpha}}\sum_{i=1}^{N_{\alpha}}
\langle\left|\bm{r}_{i}(t+t_{0}) - \bm{r}_{i}(t_{0})\right|^2\rangle,
\end{eqnarray}
with $\bm{r}_{i}(t)$ the position of particle $i$ of species $\alpha$
at time $t$, $t_{0}$ the time origin, and $N_{\alpha}$ the number
of particles of species $\alpha$.  The angular brackets correspond
to an ensemble average over the different samples.

Figure \ref{fig4} shows the MSD of A particles in $z$ direction,
i.e.~in the neutral direction perpendicular to the direction of
shear, for the supercooled liquid at $\rho = 1.2$, $T = 0.44$, and
different shear rates.  Also included in the figure is the corresponding
MSD for the unsheared liquid at equilibrium. This MSD displays the
well-known behavior for a supercooled liquid, with a ballistic
regime $\propto t^2$ at very short times, a diffusive behavior
$\propto t$ in the long-time limit, and at intermediate times a
subdiffusive plateau-like regime. Note that the MSDs for B particles
exhibit a similar behavior. The MSDs corresponding to shear rates in 
the Newtonian regime coincide with the one at equilibrium, as expected.
At higher shear rates ($\dot{\gamma}= 10^{-4}$ and $10^{-3}$ in the figure),
the MSDs are on top of the one at equilibrium up to the time where
the overshoot in the corresponding stress-strain relation occurs.
As can be seen in the figure, the stress drop in the stress-strain 
relation from $\sigma_{\rm max}$ to $\sigma_{\rm ss}$ corresponds to
a superlinear increase of the MSD towards a linear diffusive regime
when the steady state is reached.

%%%%%%%%%%%%%%%%%%%%%%%%%%%%%%%%%%%%%%%%%%%%
\begin{figure}%[!htbp]
\begin{center}
\includegraphics[width=6cm]{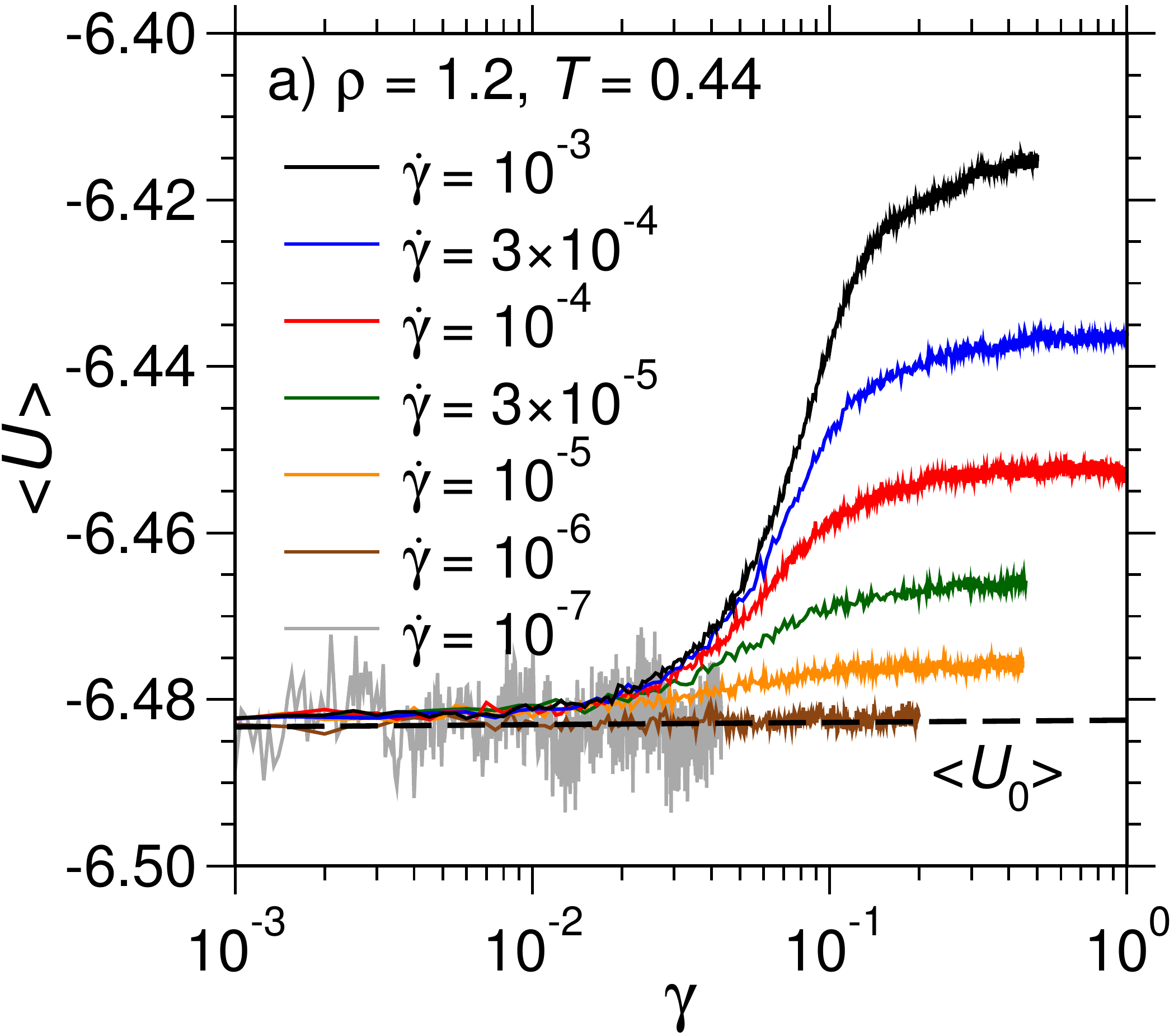}
\includegraphics[width=6cm]{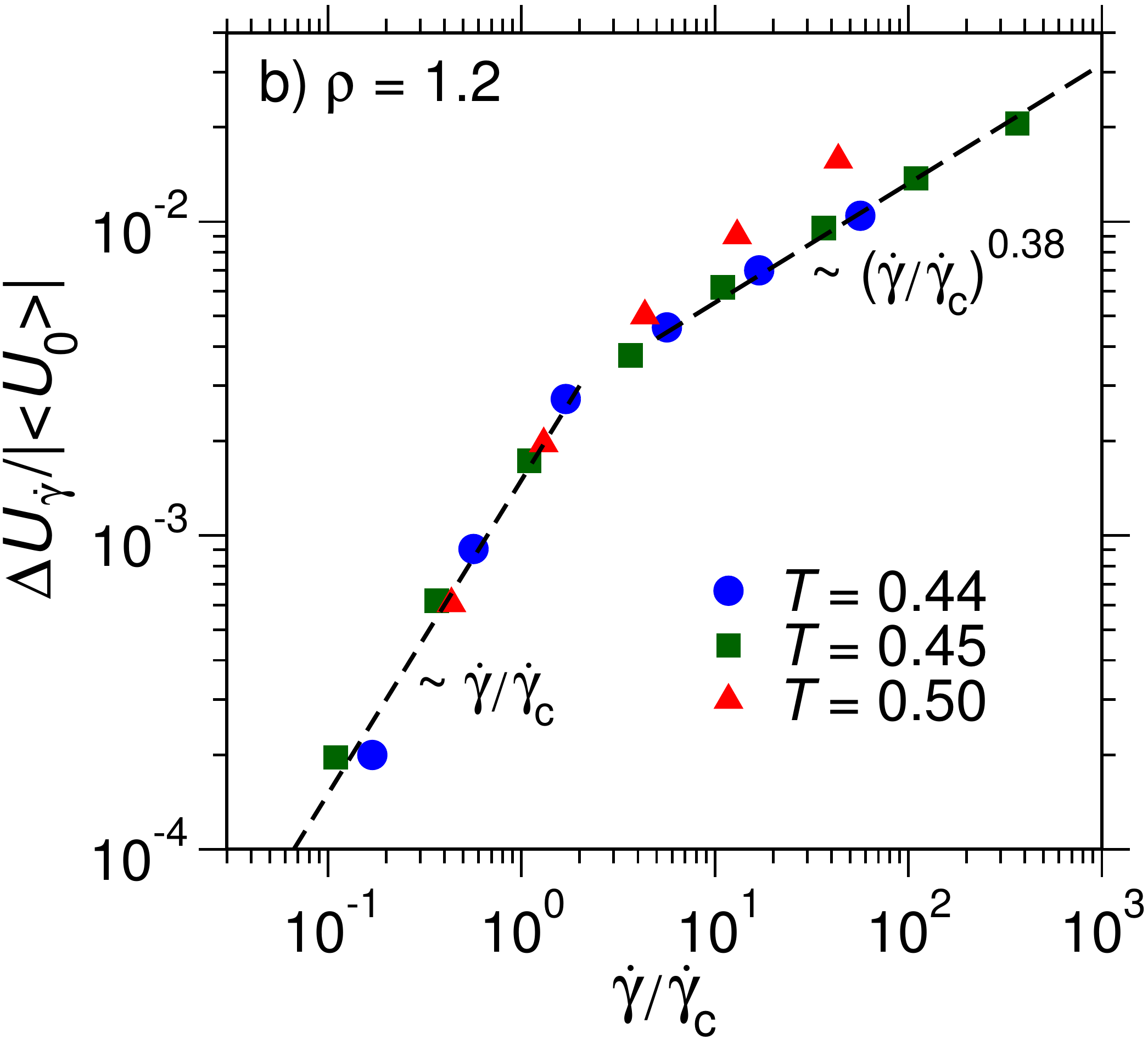}
\caption{\label{fig5} a) The potential energy, $\langle U \rangle$,
as a function of strain for the supercooled liquid at $\rho = 1.2$
and $T = 0.44$. The shear rates are $\dot{\gamma} = 10^{-3}$,
$3\times 10^{-4}$, $10^{-4}$, $3\times 10^{-5}$, $10^{-5}$, $10^{-6}$,
$10^{-7}$. The dashed line marks the potential energy of the
undeformed system at $\langle U_0 \rangle = -6.482$. b) Double-logarithmic
plot of the potential energy change $\Delta U_{\dot\gamma}/|\langle
U_0 \rangle|$ as a function of $\dot\gamma/\dot{\gamma}_c$. The
dashed lines are fits with power laws, as indicated.}
\end{center}
\end{figure}
%%%%%%%%%%%%%%%%%%%%%%%%%%%%%%%%%%%%%%%%%%%%%%%

{\bf Potential energy.} The occurrence of a finite stress in the
sheared supercooled liquid is accompanied by a change of the potential
energy $\langle U \rangle$.  Figure \ref{fig5}a shows $\langle U
\rangle$ as a function of strain at $\rho = 1.2$ and $T = 0.44$ for
different shear rates.  The transition from the elastic regime to
the plastic flow regime is associated with a monotonic increase of
$\langle U \rangle$ in the strain window $10^-2< \gamma < 1.0$
towards a constant value $\langle U_{\rm ss} \rangle (\dot{\gamma})$
in the steady state.  To quantify the change of the potential energy
from the quiescent state to the steady state, we have computed the
difference $\Delta U_{\dot\gamma} = \langle U_{\rm ss} \rangle
(\dot{\gamma}) - \langle U_0 \rangle$ (with $\langle U_0 \rangle$
being the potential energy of the unsheared liquid).  In Fig.~\ref{fig5}b,
we plot $\Delta U_{\dot\gamma}/|\langle U_0 \rangle|$ as a function
of $\dot\gamma/\dot{\gamma}_c$.  As the figure indicates, $\Delta
U_{\dot\gamma}$ changes its behavior around $\dot{\gamma}_c$,
i.e.~around $\dot\gamma/\dot{\gamma}_c = 1.0$ there is a crossover
from a linear dependence on $\dot\gamma/\dot{\gamma}_c$ to a power
law $\propto (\dot\gamma/\dot{\gamma}_c)^\alpha$ with $\alpha<1.0$.
For the two lower temperatures, $T=0.44$ and $T=0.45$, the data can
be well described by a power law with $\alpha=0.38$. This power law
can be seen as the analog to the Herschel-Bulkley law with which
we have described the flow curves in the non-Newtonian regime in
Fig.~\ref{fig3}a.

%%%%%%%%%%%%%%%%%%%%%%%%%%%%%%%%%%%%%%%%%%%%
\begin{figure*}%[!htbp]
\begin{center}
\includegraphics[width=17cm]{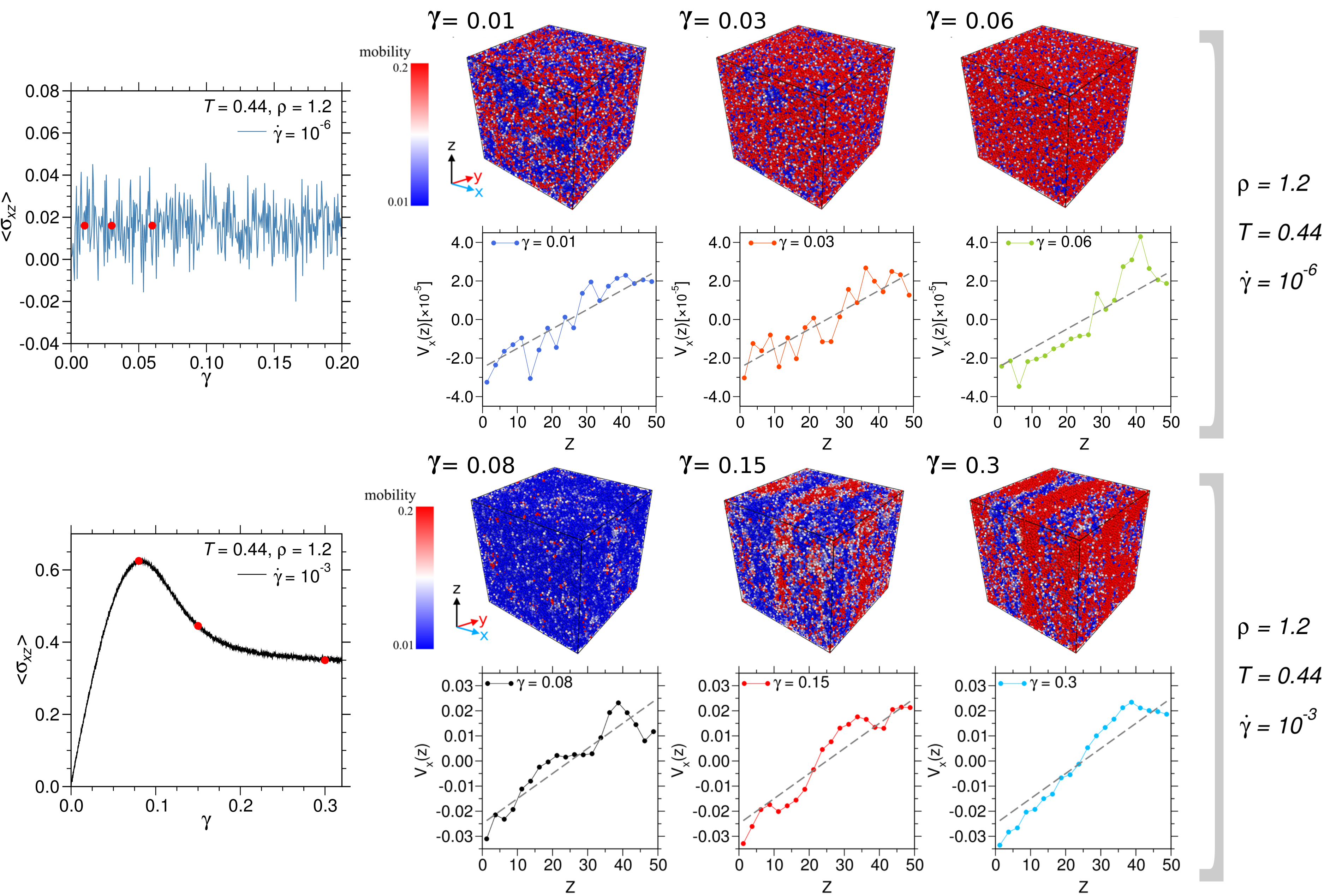}
\caption{\label{fig6} {\bf Upper panels:} Stress-strain relation,
spatial maps of the squared displacement and the corresponding
velocity profiles for the supercooled liquid at $\rho = 1.2$, $T =
0.44$, and $\dot{\gamma} = 10^{-6}$ for ${\gamma} = 0.01$, 0.03,
and 0.06 (these values of $\gamma$ are marked in the plot for
$\langle \sigma_{xz} \rangle (\gamma)$ by solid red circles). {\bf
Lower panels}: Same as the upper panel but now at $\dot{\gamma} =
10^{-3}$ for ${\gamma} = 0.08$, 0.15, and 0.3.  For both shear
rates, the linear size of the cubic samples is $L=50$.}
\end{center}
\end{figure*}
%%%%%%%%%%%%%%%%%%%%%%%%%%%%%%%%%%%%%%%%%%%%%%

{\bf Flow patterns.} Up to now, we have only considered macroscopic
properties of the supercooled liquids under shear. As a result, we
have characterized the crossover from Newtonian to non-Newtonian
behavior for the various quantities as a function of shear rate.
Now, we examine the possibility of inhomogeneous flow patterns in
the supercooled liquids due to the external shear. To this end, we
compute mobility color maps of single-particle, i.e.~non-averaged,
squared displacements $\delta r_{\alpha, z}^{2}(t)$ at a given
strain value $\gamma = \dot{\gamma} t$. Snapshots of such maps for
different values of $\gamma$ are displayed in Fig.~\ref{fig6} at
$\rho=1.2$ and $T=0.44$ for the shear rates $\dot{\gamma}=10^{-6}$
(Newtonian regime) and $\dot{\gamma}=10^{-3}$ (non-Newtonian regime).
Also included in the figure are the corresponding velocity profiles
$V_x(z)$.  The color code is chosen such that blue corresponds to
a low and red to a high squared displacement. At the low shear rate,
some small spatial heterogeneities can be seen at $\gamma=0.01$,
but already at $\gamma = 0.06$ the flow is very homogeneous with
only small spots of immobile regions.  This can be also inferred
from the velocity profiles where the deviations from the expected
linear behavior (dashed lines) can be refered to thermal noise.
Thus, one may conclude from the mobility color maps that there are
no pronounced inhomogeneous flow patterns in the Newtonian regime,
as expected.

The behavior is qualitatively different at the higher shear rate,
$\dot{\gamma} = 10^{-3}$, i.e.~in the non-Newtonian regime.  At a
strain of $\gamma = 0.08$, most of the particles have a very low
squared displacement which indicates the presence of an elastic regime
in this case, with an essentially affine deformation of the sample. At
the strain $\gamma = 0.15$, i.e.~slightly after the emergence of the
stress overshoot in the stress-strain relation (cf.~Fig.~\ref{fig1}a),
in the direction perpendicular to the shear ($x$ direction), bands of
high mobility occur while the rest of the system is still immobile. These
vertical shear bands are still present at $\gamma = 0.3$.  The vertical
bands can be, of course, not clearly identified in the corresponding
velocity profiles, because they are only sensitive to flow patterns in
the direction of shear. 

We have identified a critical shear rate $\dot{\gamma}_c$ which
marks the crossover of the response of the supercooled liquid to
the external shear from Newtonian to non-Newtonian behavior.
Characteristic features of the non-Newtonian regime are the occurrence
of an overshoot in the stress-strain relation, an interim
Herschel-Bulkley-like behavior in the flow curve and the potential
energy, and a superlinear regime prior to the diffusive steady-state
regime in the mean-squared displacement. Furthermore, we observe
short-lived vertical shear bands in the non-Newtonian regime that,
as we shall see below, also encounter in glasses under shear.

\subsection{Glasses under shear}
In this section, we analyze glasses under shear. Here, we consider
states at the density $\rho=1.3$ and the temperature $T=10^{-4}$.
The choice of temperature is such that there are finite but very
small thermal fluctuations, but the affine deformation due to the
external drive would dominate.  Unlike the supercooled liquids, the
response of the glass states to the external shear field differs
significantly from sample to sample.  Therefore, most of the
properties that are shown in the following are not obtained by an
average over many samples, but we discuss them for individual
samples. In particular, even inhomogeneous flow patterns are
qualitatively different from sample to sample.

%%%%%%%%%%%%%%%%%%%%%%%%%%%%%%%%%%%%%%%%%%%%
\begin{figure*}[!htbp]
\begin{center}
\includegraphics[width=15cm]{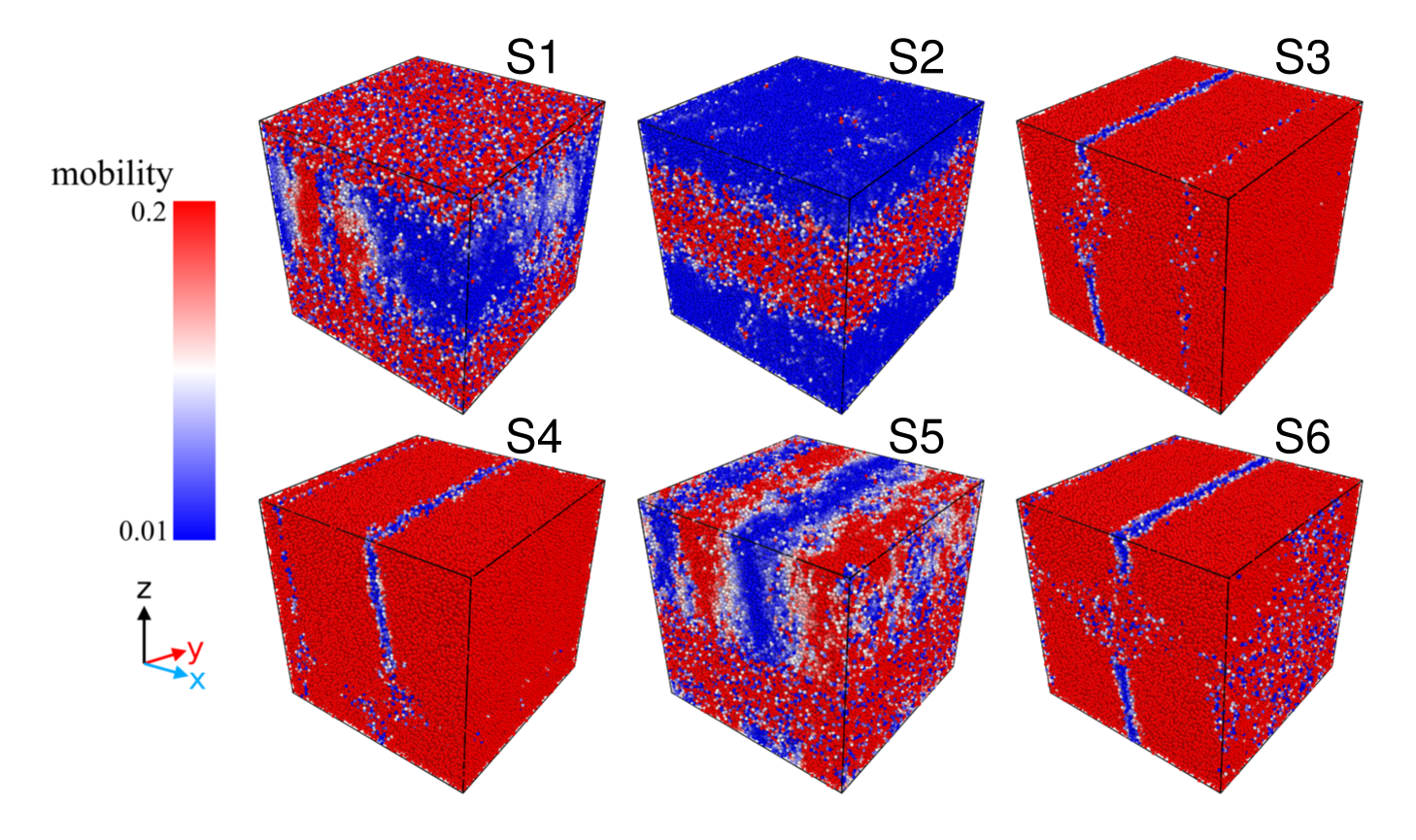}
\caption{\label{fig7} Snapshots of spatial maps of the squared
displacement, using different initial glass states, but all with
same thermal history, at the strain $\gamma = 1.0$. The density,
temperature, and shear rate are $\rho = 1.3$, $T = 10^{-4}$, and
$\dot{\gamma} = 10^{-4}$, respectively.  The linear size of the
cubic samples is $L = 60$.}
\end{center}
\end{figure*}
%%%%%%%%%%%%%%%%%%%%%%%%%%%%%%%%%%%%%%%%%%%%%%%

{\bf Vertical and horizontal shear bands.} This is illustrated by
the snapshots of spatial maps of the squared displacement at the
strain $\gamma = 1.0$ in Fig.~\ref{fig7}.  The corresponding initial
configurations before the switch-on of the shear are cubic glass
samples with linear dimension $L=60$ at $\rho=1.3$ and $T=10^{-4}$
with a similar thermal history (see above). However, the flow
patterns that one can see in the snapshots are obviously different.
One can identify two types of shear bands.  While the first and
second sample has a horizontal shear band, the third, fourth and
sixth sample display two vertical shear bands. A mixed situation
where both types of shear bands are present can be observed in the
fifth sample.  We note that at the considered fixed strain of $\gamma
= 1.0$, the width of the horizontal bands corresponds to about a
quarter of the linear dimension of the simulation box ($L=60$).  In
the cases where two vertical shear bands form after the stress
overshoot, these bands have almost merged at $\gamma=1.0$ and the
mobility map essentially shows a homogeneously flowing fluid in the
steady state.

%%%%%%%%%%%%%%%%%%%%%%%%%%%%%%%%%%%%%%%%%%%%%%%%%%%%%%%%%%%%%%
\begin{figure}[!htbp]
\begin{center}
\includegraphics[width=4.2cm]{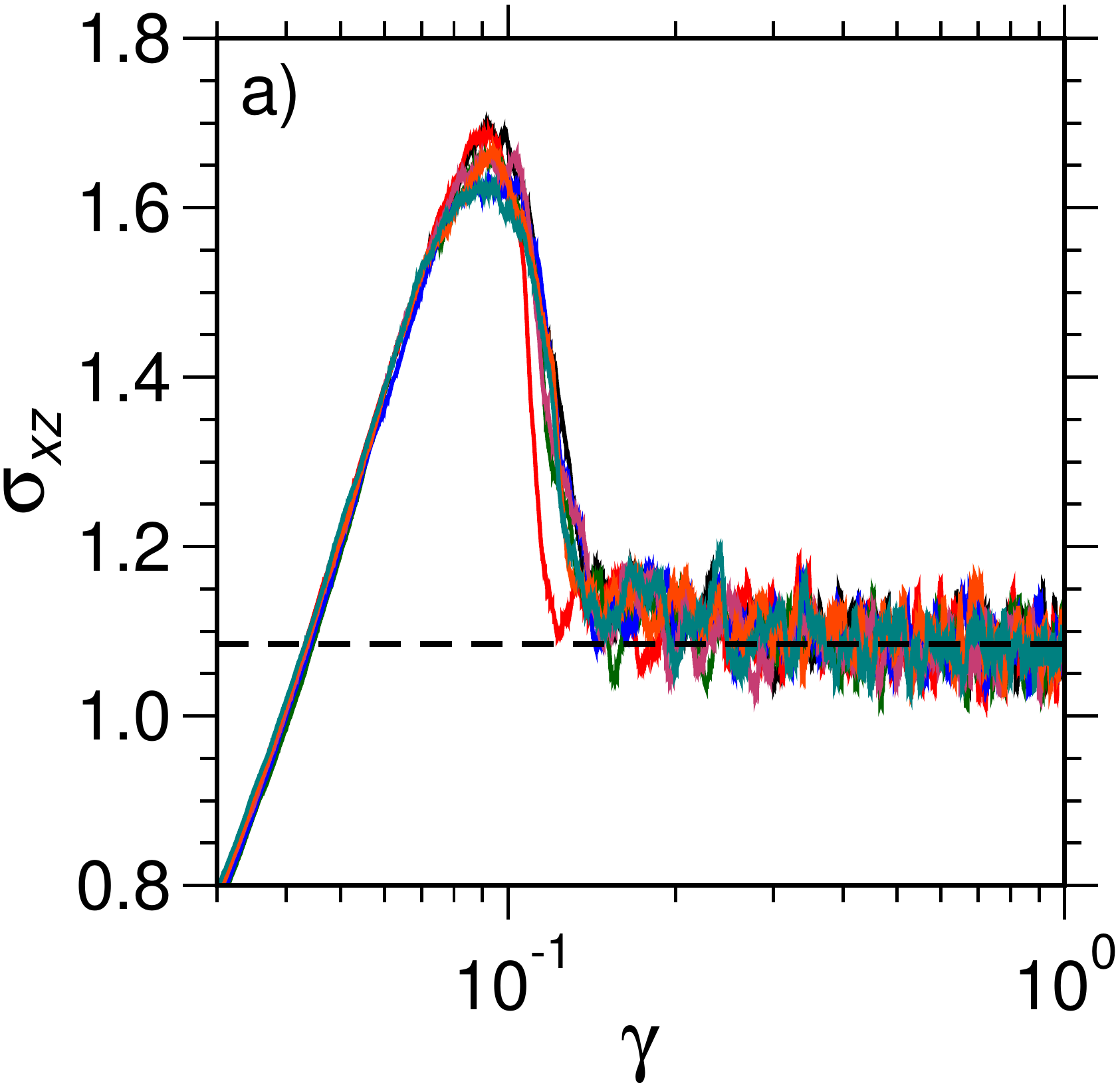}
\includegraphics[width=4.2cm]{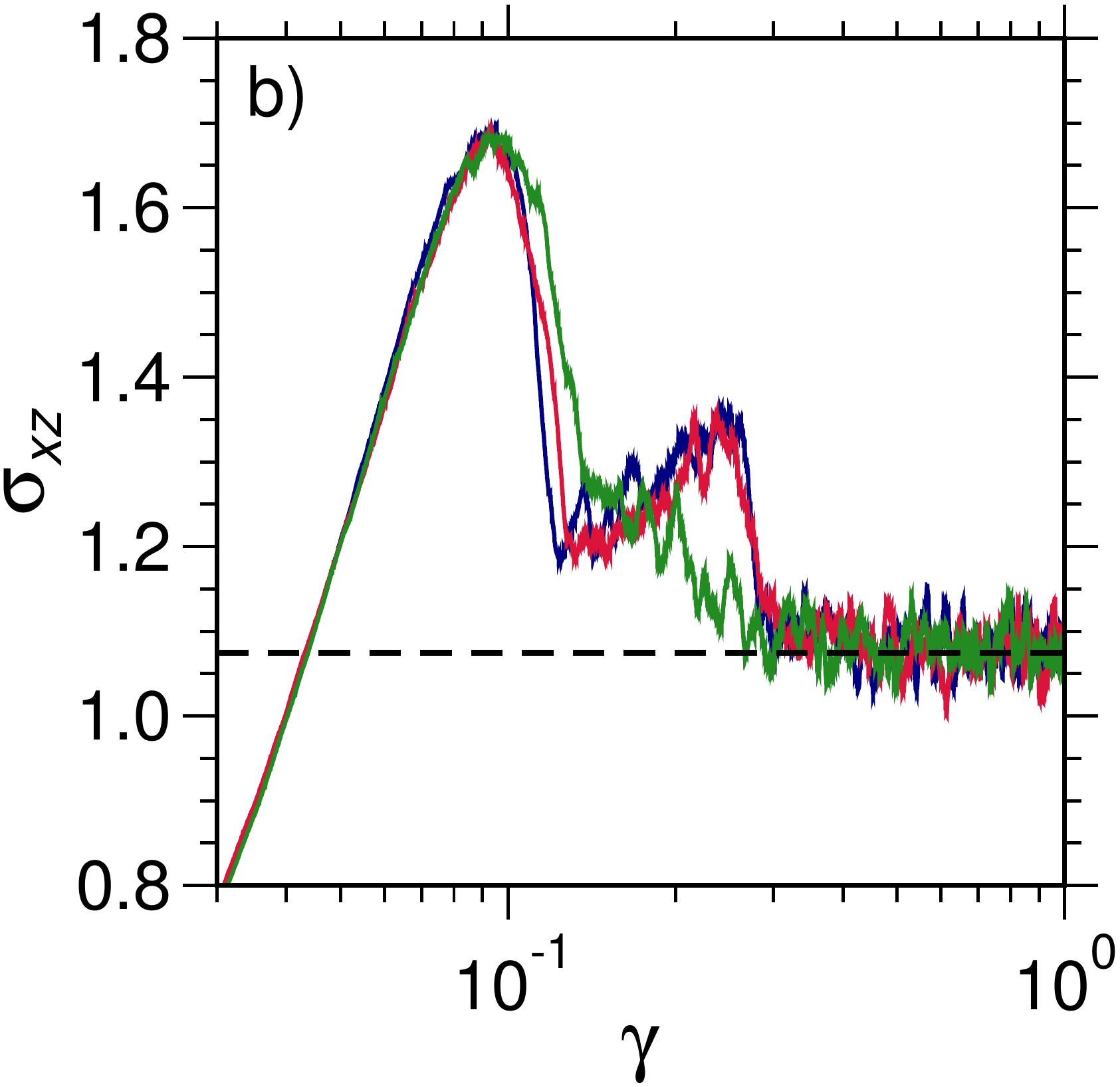}
\includegraphics[width=4.2cm]{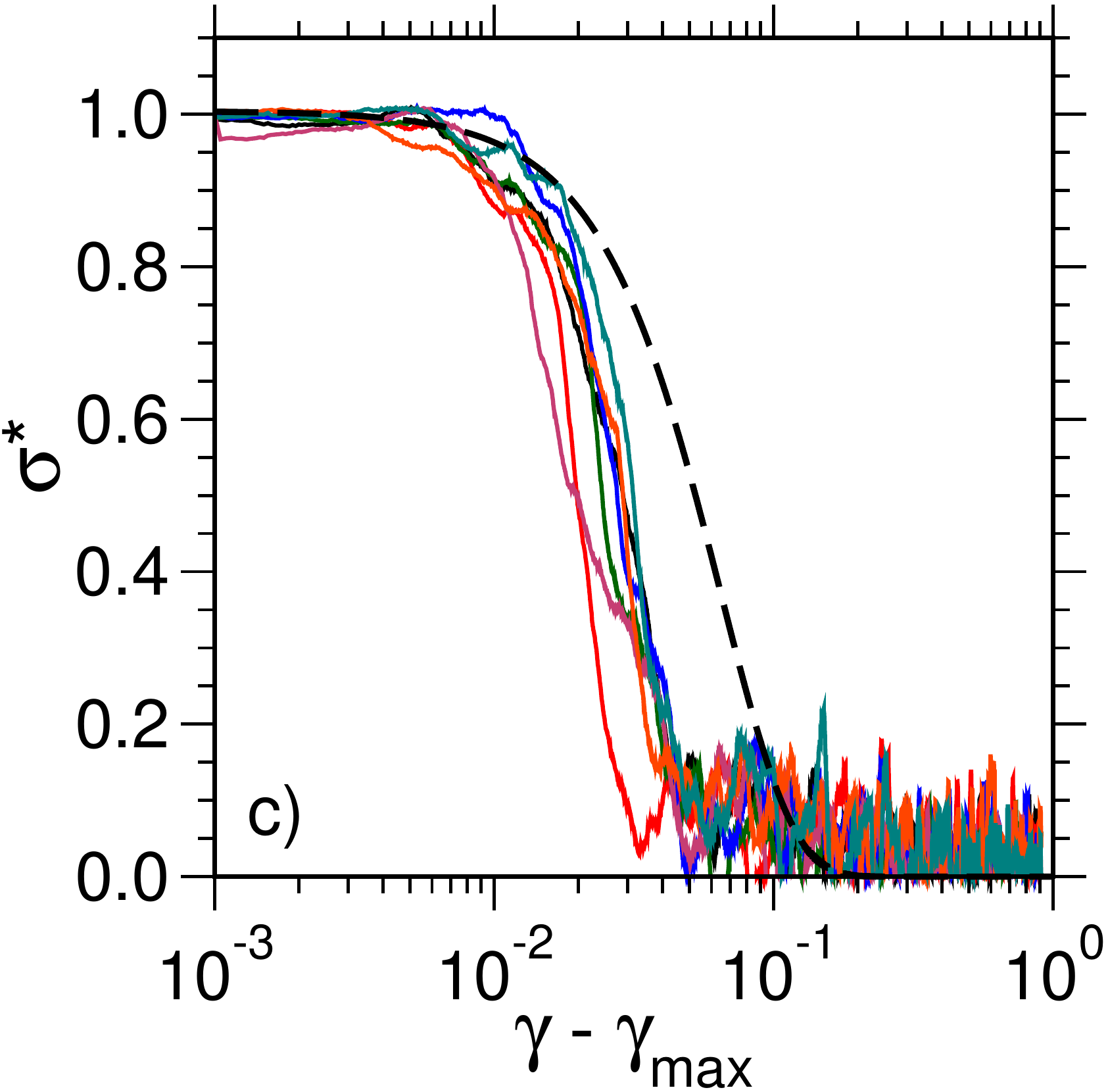}
\includegraphics[width=4.2cm]{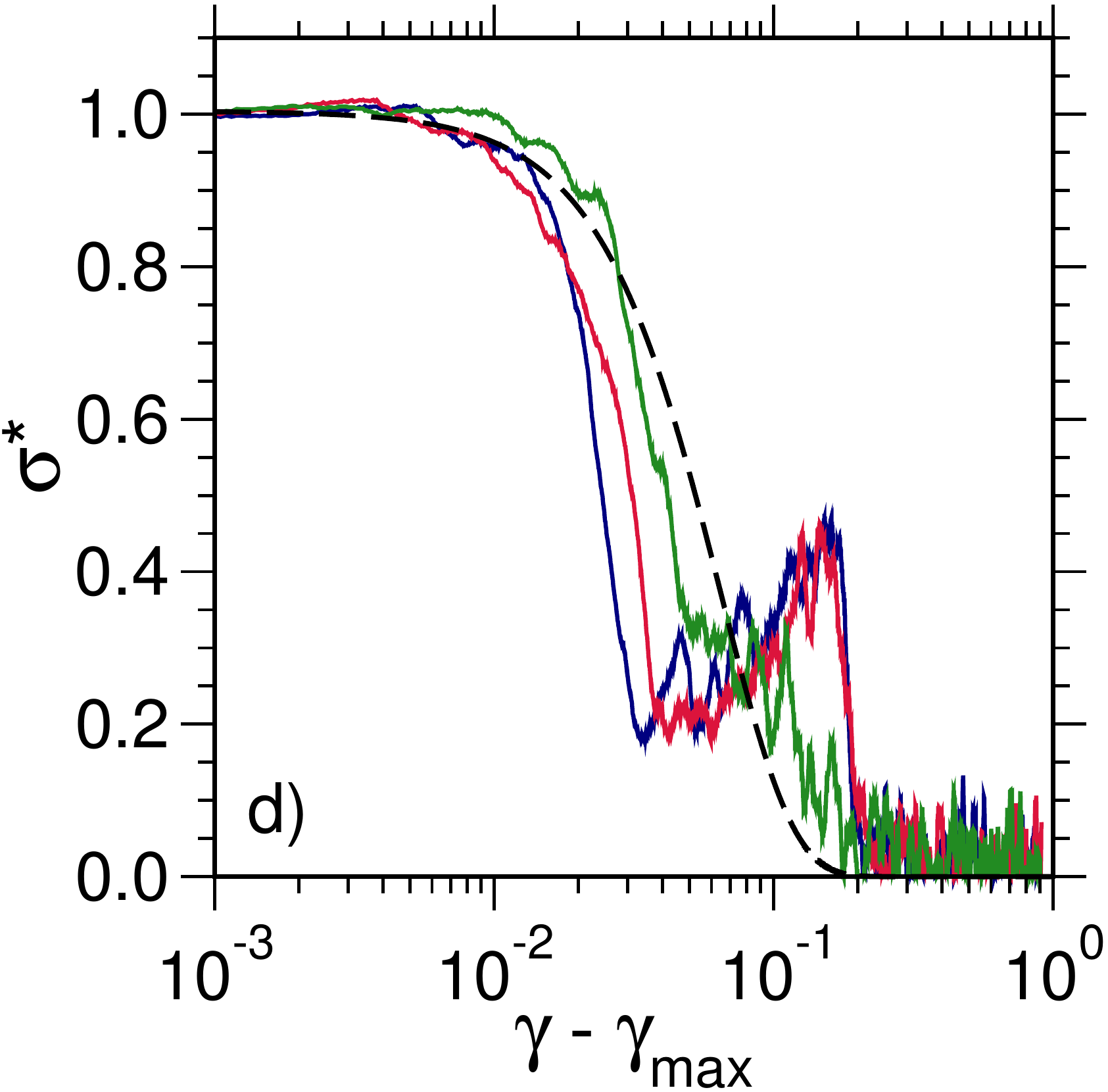}
\includegraphics[width=4.2cm]{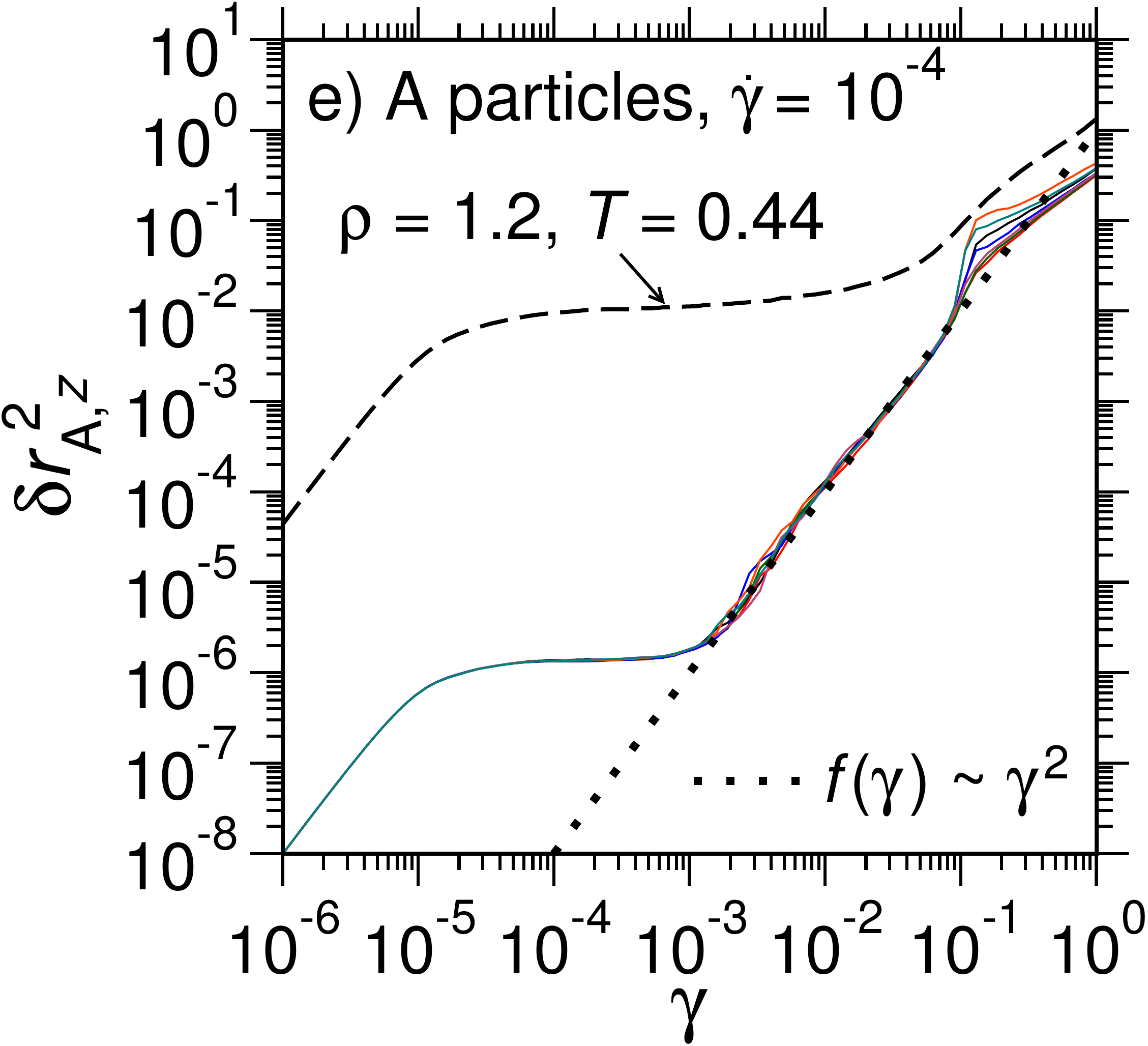}
\includegraphics[width=4.2cm]{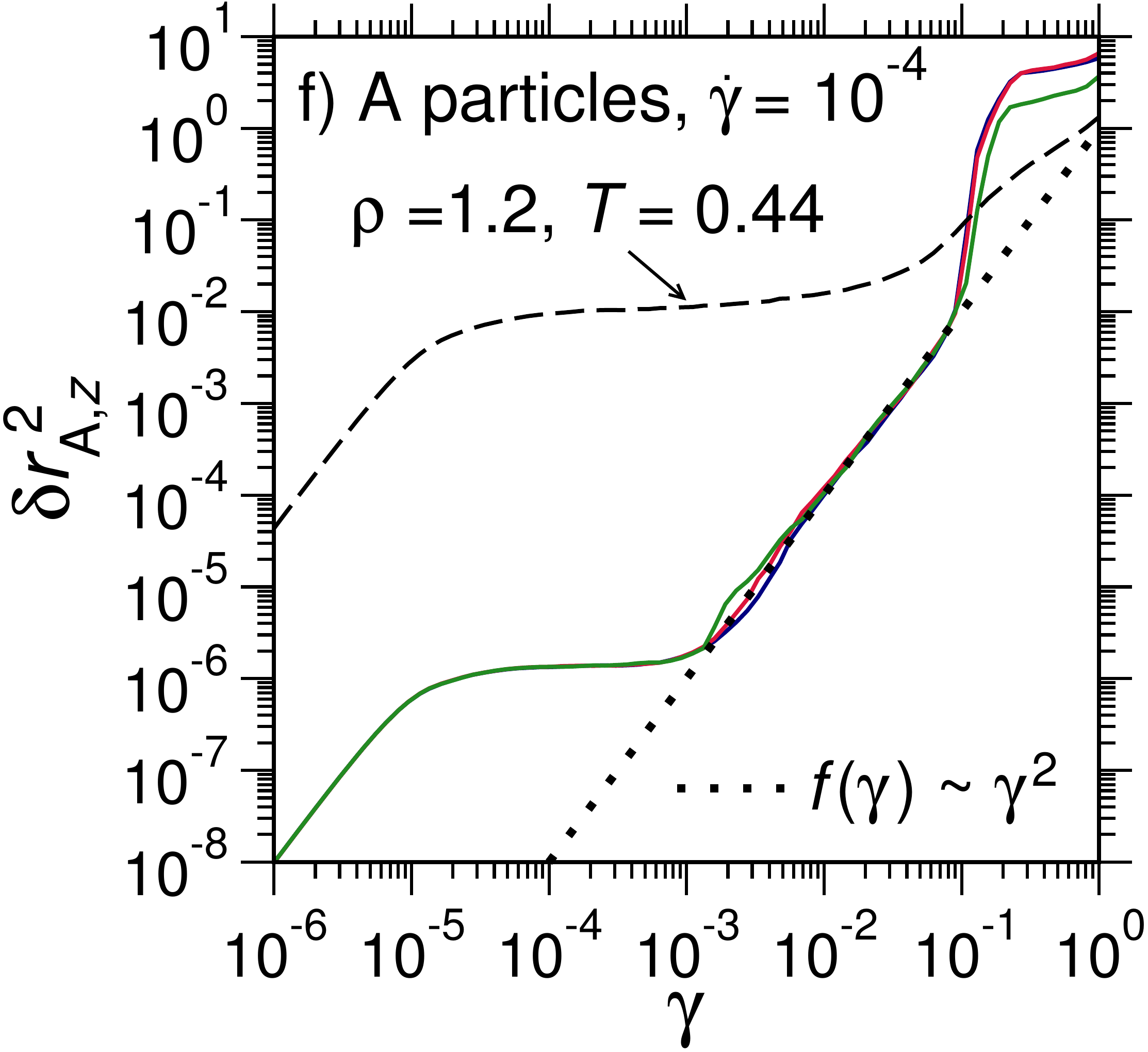}
\caption{\label{fig8} a) and b) Stress-strain relations at $\rho =
1.3$, $T = 10^{-4}$, and $\dot{\gamma}=10^{-4}$ for different
individual cubic samples with linear size $L=60$. The horizontal
dashed lines mark the steady-state stress $\sigma_{\rm ss}$. c) and
d) Reduced stress $\sigma^{\star}$ as a function of $\gamma -
\gamma_{\rm max}$, corresponding to the data shown in a) and b),
respectively.  The dashed black lines show the compressed exponential
function from Fig.~\ref{fig1}b.  e) and f) The $z$ component of the
squared displacement for A particles, $\delta r_{{\rm A}, z}^2$,
as a function of $\gamma$ for different samples, corresponding to
those used in a) and b), respectively (in e) and f), the result is
averaged over all the A particles in each sample).  In e) and f),
the dotted lines represent the function $f(\gamma)=a^2 \gamma^2$
with $a=1.0\,\sigma_{\rm AA}$.  Also the MSD for the supercooled
liquid at $\rho=1.2$, $T=0.44$, and $\dot{\gamma}=10^{-4}$ is
included in e) and f).}
\end{center}
\end{figure}
%%%%%%%%%%%%%%%%%%%%%%%%%%%%%%%%%%%%%%%%%%%%%%%%%%%

The stress-strain relations in Figs.~\ref{fig8}a and \ref{fig8}b
correspond respectively to cases with horizontal and vertical shear
bands.  Obviously, the stress-strain relations are qualitatively
different in both cases.  In the case of horizontal shear bands
(Fig.~\ref{fig8}a), the stress has a maximum at $\gamma_{\rm max}
\approx 0.09$, followed by a relatively sharp drop and subsequently
a weakly decreasing function towards the steady-state stress. Note
that the shear band forms just after the stress drop, as we discuss
below. In the case of the vertical shear bands (Fig.~\ref{fig8}b),
there is a second peak or at least a shoulder after the main stress
overshoot. Here, typically two vertical bands are formed after the
first overshoot.  The fact, that the stress tends to increase again
after the occurrence of the first stress drop, indicates that
compared to the case of horizontal bands, the vertical ones are
rather unstable with a short lifetime and indeed this is what we
observe in our simulations. We note that a similar behavior of the
stress-strain relation in case of vertical shear bands, i.e.~an
increase of the stress after the stress overshoot, has been also
found in a recent simulation using an aqs protocol \cite{kapteijns2019,
ozawa2019}. While the horizontal shear bands are associated with a
larger initial stress drop, in the case of vertical bands, the
system evolves much faster towards the steady state. In the case
of horizontal bands, the initial stress release is followed by a
relatively slow broadening of the shear band (see below).

In Figs.~\ref{fig8}c and \ref{fig8}d, the reduced stress $\sigma^\star$
as a function of $\gamma-\gamma_{\rm max}$ for the samples corresponding
respectively to those in Figs.~\ref{fig8}a and \ref{fig8}b are
shown. Also included in these plots is the compressed exponential
function with which we have fitted the data for the supercooled
liquids in Fig.~\ref{fig1}b.  In the case of the horizontal shear
bands, there is a first decay which is clearly faster than that
observed for the superooled liquid. However, this fast drop is
followed by a slowly decaying tail which is associated with the
broadening of the horizontal shear band. Unfortunately, the quality
of the data does not allow to analyze the functional behavior of
the latter tail.  Also in the case of the vertical bands
(Fig.~\ref{fig8}d), the first stress drop tends to be faster than
in the case of the supercooled liquids. Then, there is the occurrence
of a second peak or shoulder before quickly approaching the steady
state for $\gamma<1.0$.

The $z$ component of the squared displacement for A particles,
$\delta r_{{\rm A}, z}^2$, as a function of $\gamma$ for the samples
with horizontal and vertical shear bands are plotted in Fig.~\ref{fig8}e
and \ref{fig8}f, respectively.  In both plots, we have also included
for comparison, the corresponding MSD for the supercooled liquid
at $\rho=1.2$, $T=0.44$, and $\dot{\gamma}=10^{-4}$, which reveals
a completely different behavior of the squared displacement for the
glass states at $T=10^{-4}$, as we discuss now. For $\gamma< 10^{-3}$,
an expected behavior is seen, i.e. after an initial ballistic regime,
a plateau at the value $\delta r_{{\rm plat}}^2\approx1.3\times
10^{-6}$ is reached which reflects the strong localization of the
particle at the extremely low temperature $T=10^{-4}$ (note that a
similar value is obtained for the plateau value of the corresponding
unsheared glass state). However, in the interval $10^{-3} < \gamma
< 10^{-1}$, there is a ballistic regime where $\delta r_{{\rm A},
z}^2 = \gamma^2 a^2 = (u t)^2$, with $u$ a velocity and $a$ a
microscopic length scale of order 1 (in Figs.~\ref{fig8}e and
\ref{fig8}f we have chosen $a=1.0$).  This regime sets in when the
condition $\gamma a \approx \sqrt{\delta r_{{\rm plat}}^2}$ holds,
i.e.~at about $\gamma \approx 10^{-3}$.  Then, the deformation due
to the shear dominates and the small thermal fluctuations are not
relevant anymore. However, the strain is not yet sufficient to break
the cage around the tagged particle, thus inducing a plastic flow
event. This requires still a strain of the order of 0.1. Around the
latter strain, plastic flow sets in, which is associated with
horizontal shear bands (Fig.~\ref{fig8}e) or vertical shear bands
(Fig.~\ref{fig8}f). As the figures show, for the cases with vertical
bands there is a jump in the squared displacements around $\gamma=0.1$
which is much more pronounced than in the case of the horizontal
bands.

%%%%%%%%%%%%%%%%%%%%%%%%%%%%%%%%%%%%%%%%%%%%%%%%%%%%%%%%%%%%%%
\begin{figure}[!htbp]
\begin{center}
\includegraphics[width=7.cm]{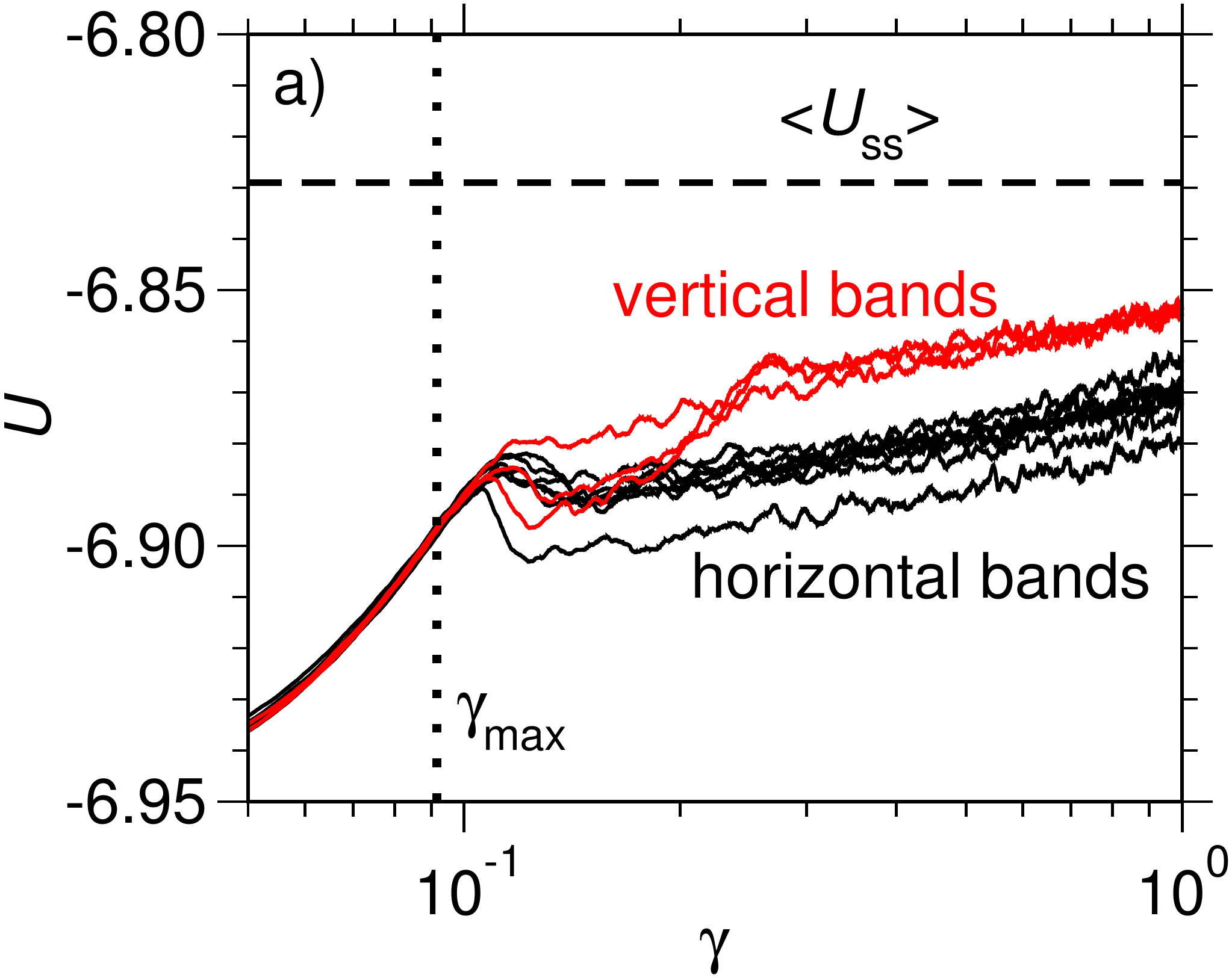}
\includegraphics[width=7.2cm]{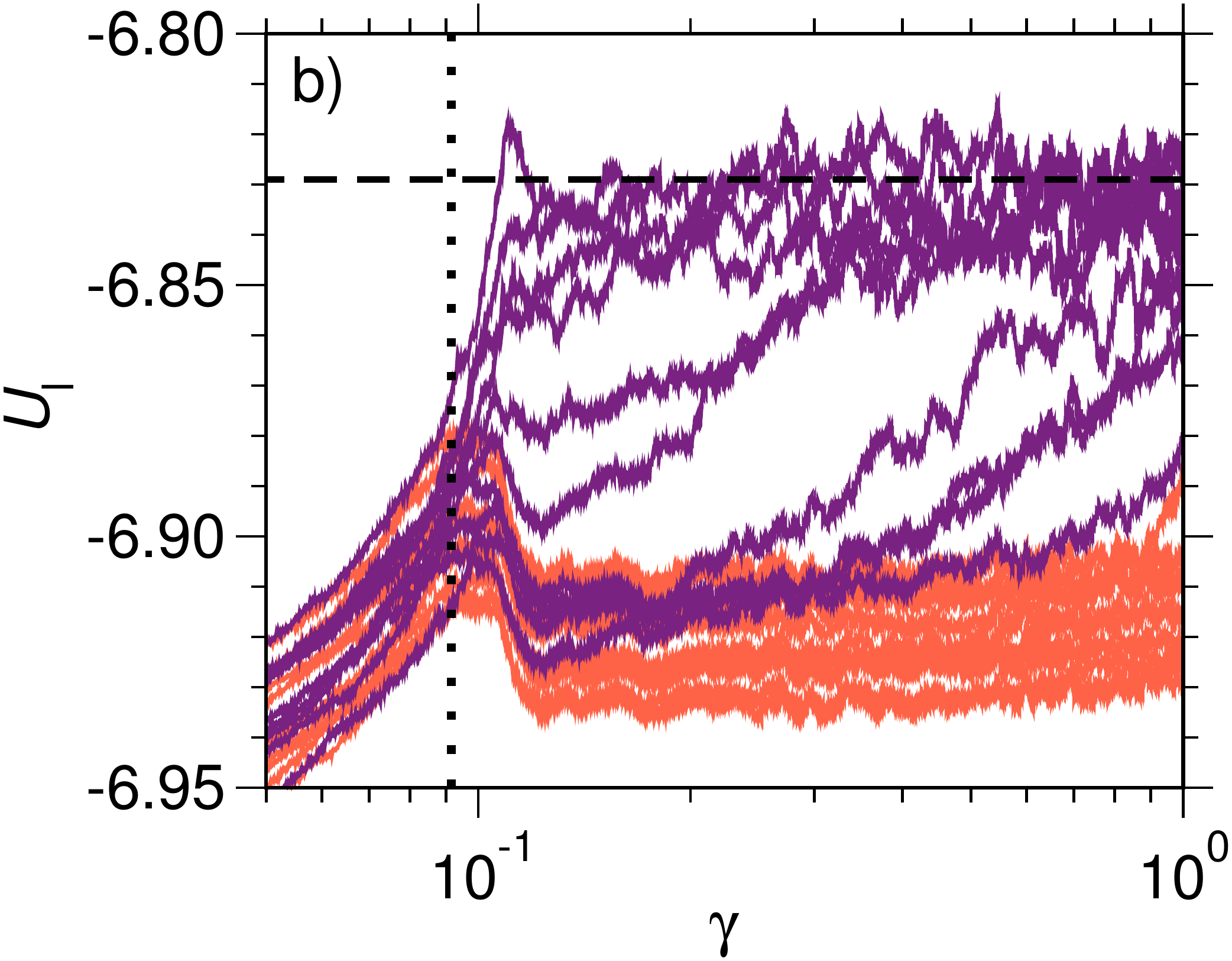}
\caption{\label{fig9} a) Potential energy $U$ as a function of
strain $\gamma$ for different samples with horizontal shear band
(black solid line) and vertical shear band (red solid line) at $\rho
= 1.3$, $T=10^{-4}$, and $\dot{\gamma} = 10^{-4}$. b) Potential
energy, $U_{\rm l}$, for different layers of a sample with horizontal
shear band as a function of strain.  The linear size of the cubic
sample is $L = 60$ and it is divided into twenty layers along the
$z$ direction, each with a width equal to $3\sigma_{\rm AA}$. The
layers that are at $\gamma=1.0$ outside of the shear band are shown
by red solid lines while those inside the shear band at $\gamma=1.0$
are represented by the violet solid lines. The horizontal dashed
and the vertical dotted lines in both panels mark the average value
of the potential energy in the steady state $\langle U_{\rm ss}
\rangle = -6.829$ and the strain at the stress overshoot $\gamma_{\rm
max}=0.0916$, respectively.}
\end{center}
\end{figure}
%%%%%%%%%%%%%%%%%%%%%%%%%%%%%%%%%%%%%%%%%%%%%%%%%%%%%%%%%%%%%%

Very pronounced differences between vertical and horizontal shear
bands can be also seen in the strain dependence of the potential
energy (Fig.~\ref{fig9}a).  Up to $\gamma\approx 0.1$, the curves
for the different samples are on top of each other and, as is the
case for the supercooled liquids (cf.~Fig.~\ref{fig5}a), a monotonic
increase of the function $U(\gamma)$ is observed.  For $\gamma >
0.1$, i.e.~when the plastic flow regime sets in, especially in the
case of the horizontal shear bands there is an overshoot, followed
by a local minimum and a subsequent increase towards the average
steady state value of the potential energy \cite{vishwas1, vishwas2},
marked by the dashed horizontal lines at $\langle U_{\rm ss} \rangle
= -6.829$ in the two panels of Fig.~\ref{fig9}. So, when the
horizontal band is nucleated, the system jumps first to a value of
the potential energy which is significantly below the steady state
value.  This means that the formation of the horizontal band provides
a significant release of stress and a lowering of the potential
energy. In the case of the vertical bands, the potential energy
tends to increase monotonically towards the steady state value. As
we have already inferred from the stress-strain relations
(cf.~Fig.~\ref{fig8}), the formation of vertical bands leads to a
faster approach of the steady state.  However, the formation of a
horizontal band is associated with a more efficient stress release
and a lower potential energy.  Thus, the monitoring of how the
potential energy (or pressure) behaves with increasing strain can
indicate the spatial orientation of the occurring shear bands.

As we shall see below, the horizontal shear band is transient and
broadens as a function of strain. Furthermore, the system with a
horizontal shear band corresponds to a state with a local potential
energy minimum. This state is very heterogeneous with respect to
the spatial distribution of potential energy \cite{shi2007}. To
analyse this issue, we consider now a sample with horizontal shear
band (sample 2 in Fig.~\ref{fig7}).  We divide the simulation box
of this sample into 20 layers along the $z$ direction, each with a
width of $3\sigma_{AA}$, and investigate the evolution of the
potential energy, $U_{\rm l}$, in each layer.  In Fig.~\ref{fig9}b,
this quantity is shown for each of the 20 layers as a function of
$\gamma$.  Different colors are assigned to the layers that at
$\gamma=1.0$ are inside the shear band region (violet solid lines)
and those that are outside the shear band region (red solid lines).
Firstly, such spatial resolution allows us to locate the strain
value at which the flow heterogeneity sets in, viz. where the
divergence of the curves for the different layers occurs, and we
note that this happens at a strain value larger than $\gamma_{\rm
max}$. Further, the plot also clearly shows that in the shear band
the potential energy is already close to the steady-state value,
while outside the shear band the system is at significantly lower
energy of the order of $-6.92$.  The growth of the shear band
provides a homogenization of the system and eventually homogeneous
flow in the steady state.  During this homogenization, the stress
slightly decreases towards the steady-state value $\sigma_{\rm ss}$
(cf.~Fig.~\ref{fig8}a).  So the increase of the potential energy
from a minimum value after the stress drop to the steady-state value
$\sigma_{\rm ss}$, is connected with a further release of the
system's stress.

%%%%%%%%%%%%%%%%%%%%%%%%%%%%%%%%%%%%%%%%%%%%%%%%%%%%%%%%%%%%%%
\begin{figure}[!htbp]
\begin{center}
\includegraphics[width=7cm]{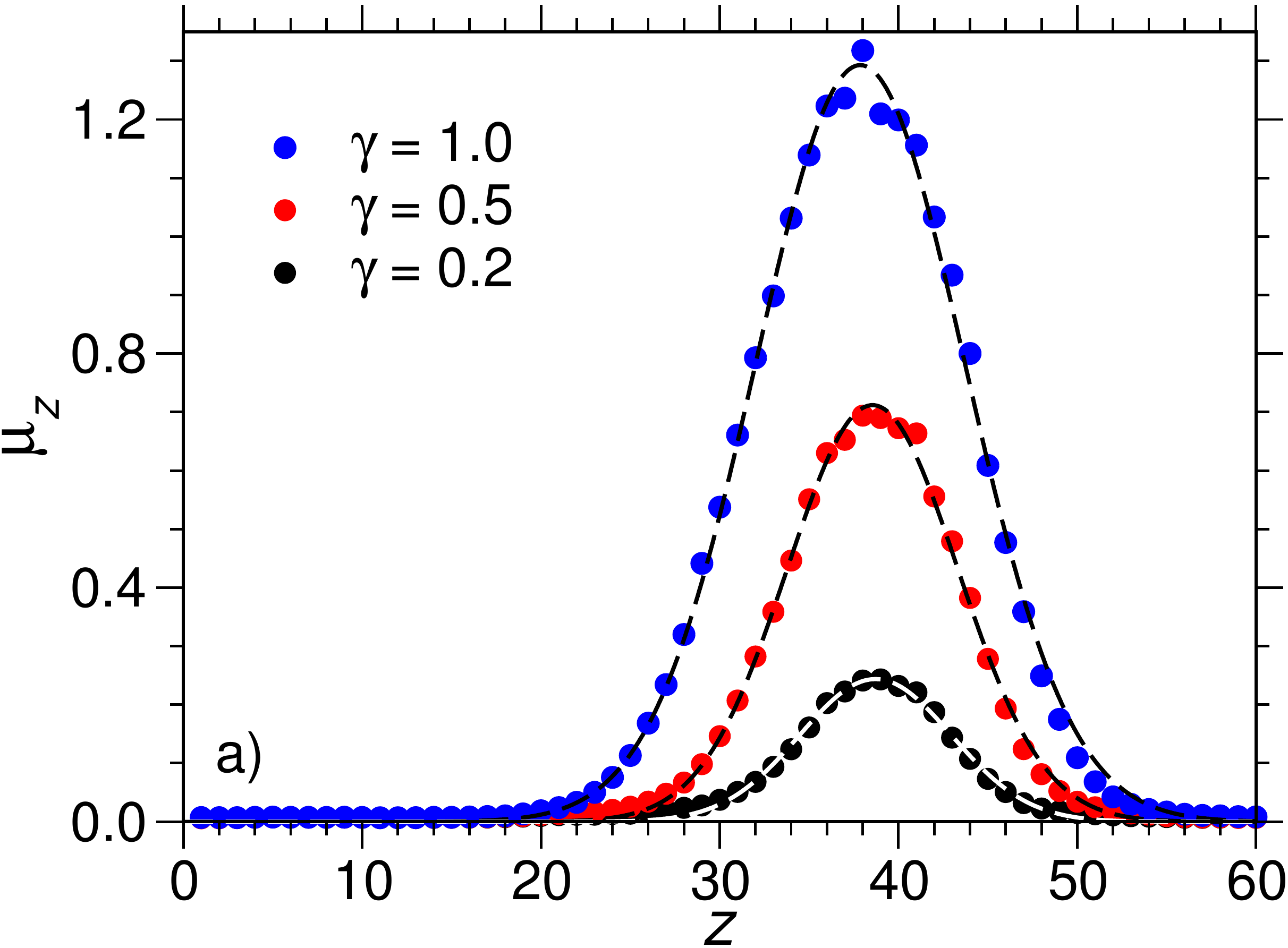}
\includegraphics[width=7.2cm]{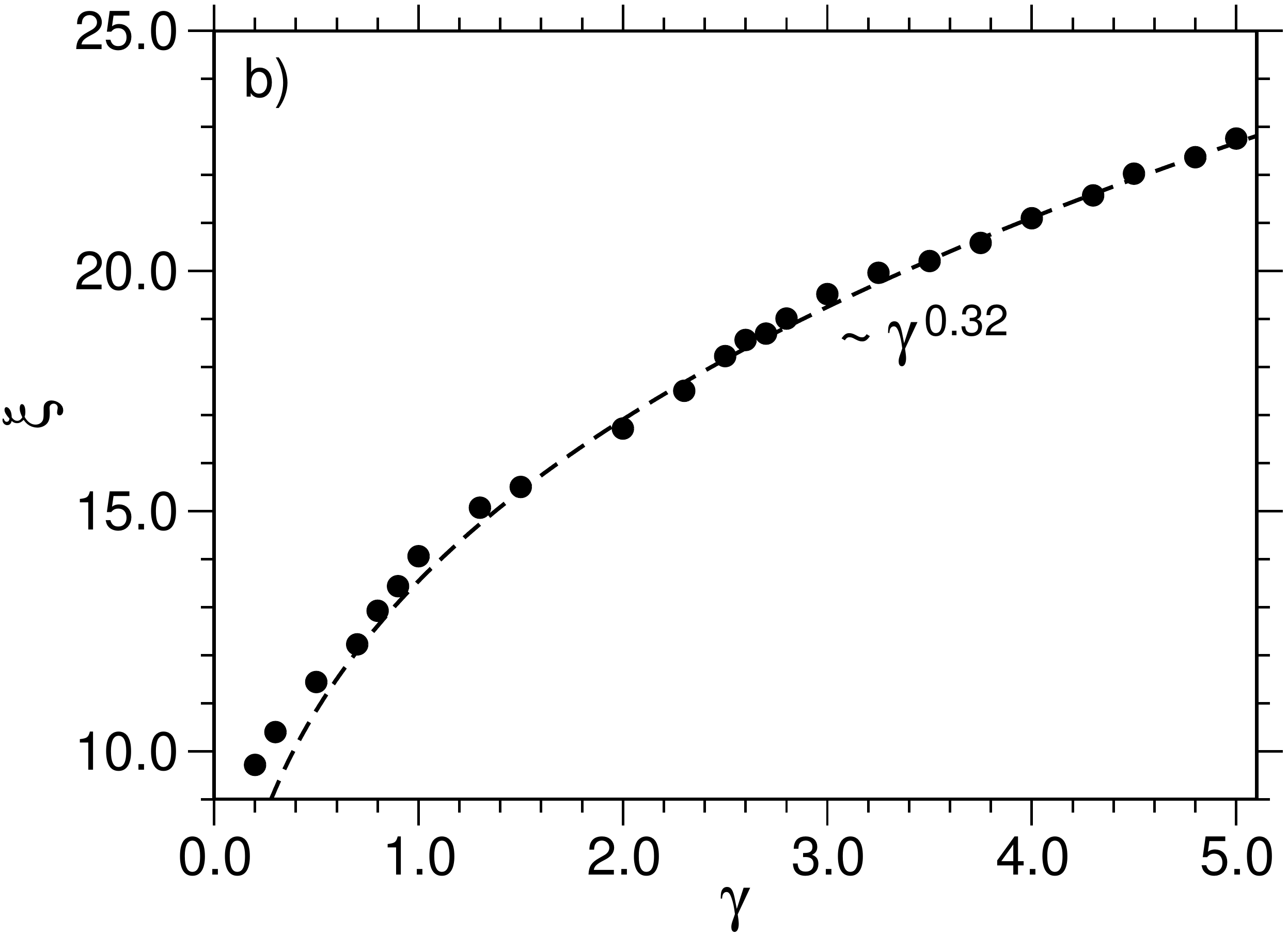}
\caption{\label{fig10} a) Mobility profiles, $\mu_{z}$, as obtained
via spatial squared displacement maps of a glass sample with
horizontal shear band at $\rho = 1.3$, $T = 10^{-4}$, and $\dot{\gamma}
= 10^{-4}$ for $\gamma = 0.2$, 0.5, and 1.0. The dashed lines are
fits with Gauss functions. b) Width $\xi$ of the shear band as a
function of strain (see text). The dashed line represents a power-law
fit with exponent $0.32$.}
\end{center}
\end{figure}
%%%%%%%%%%%%%%%%%%%%%%%%%%%%%%%%%%%%%%%%%%%%%%%%%%%%%%%%%%%%%

{\bf Growth of the shear band.} In the case of a sample with a
horizontal shear band (sample two in Fig.~\ref{fig7}), we examine
in Fig.~\ref{fig10} how the shear band grows with increasing strain.
To obtain the width of the shear-banded region, we first calculate
the mobility profiles along the $z$-direction at different strains.
The mobility profiles are defined as the average squared displacement
of particles as a function of the distance in $z$-direction. We
note that, in our case, the $z$-direction is the gradient direction.
In Fig.~\ref{fig10}a, we plot the mobility profiles at the three
different strains $\gamma = 0.2$, 0.5, and 1.0. We fitted the
mobility profiles with a Gauss function. The dependence of the width
of these Gauss functions, $\xi$, on strain is plotted in
Fig.~\ref{fig10}b. For $\gamma>1$, the data for $\xi$ can be well
described by a power law, $\xi \propto \gamma^{0.32}$, i.e.~we find
a subdiffusive growth of the horizontal shear band, in agreement
with a recent finding by Alix-Williams {\it et al} for a Cu-Zr glass
model \cite{alix2018}.

%%%%%%%%%%%%%%%%%%%%%%%%%%%%%%%%%%%%%%%%%%%%%%%%%%%%%%%%%%%%%%
\begin{figure*}[!htbp]
\begin{center}
\includegraphics[width=15cm]{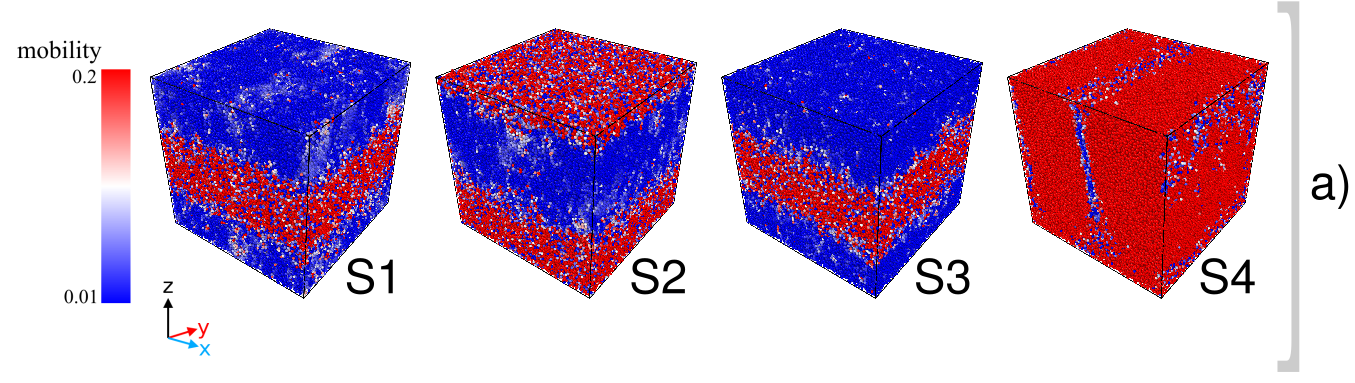}
\includegraphics[width=5cm]{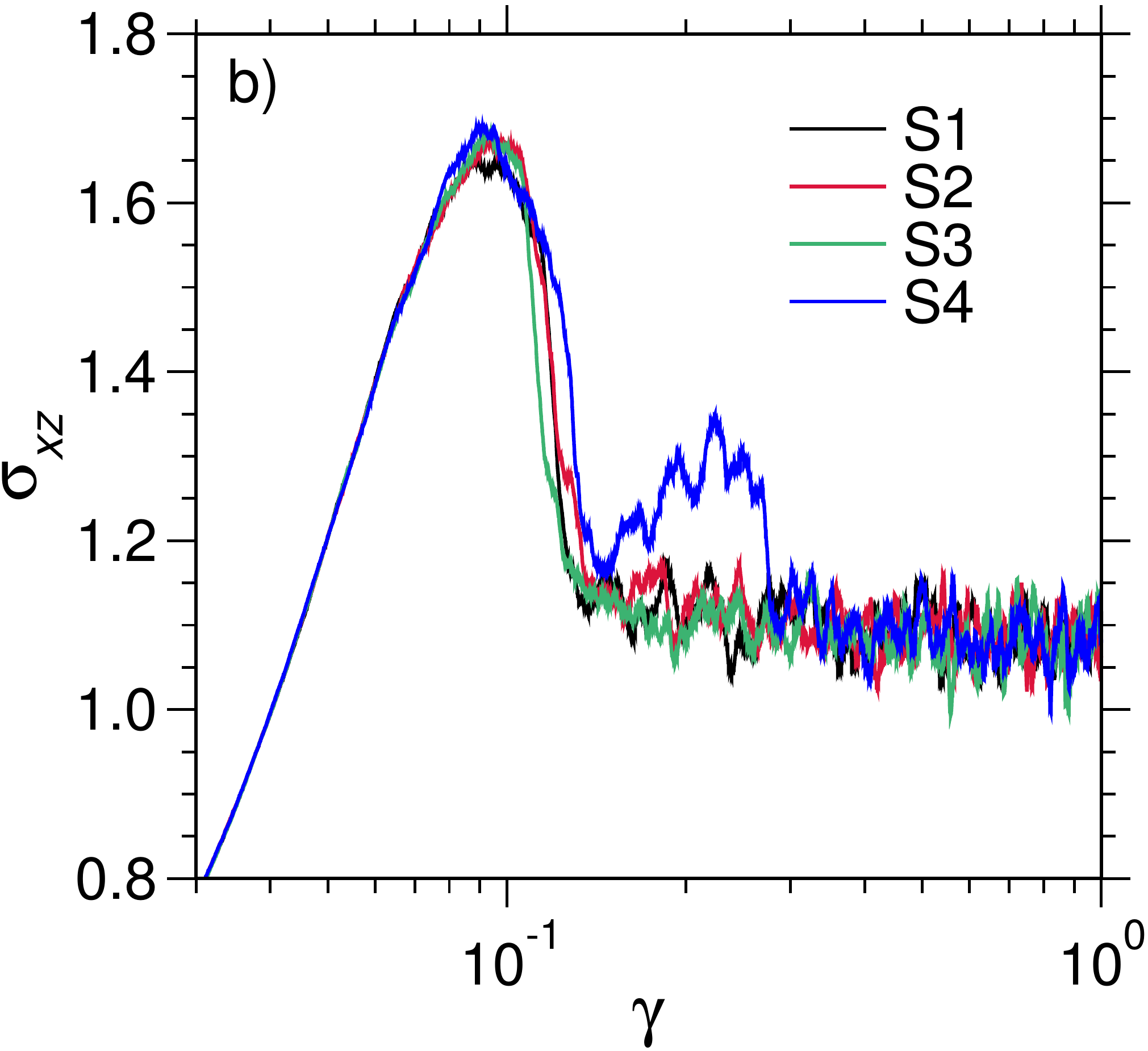}
\includegraphics[width=5cm]{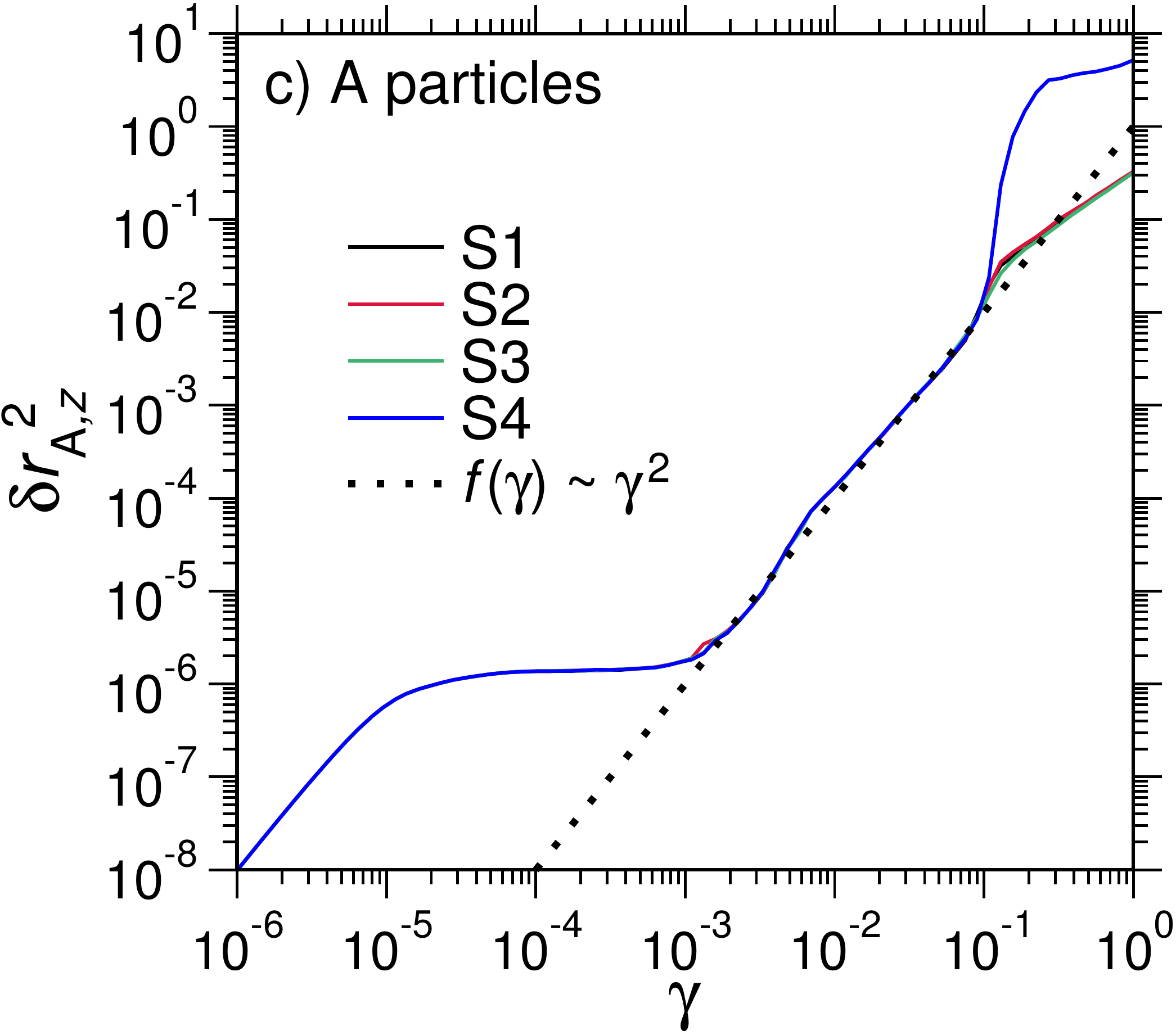}
\includegraphics[width=5cm]{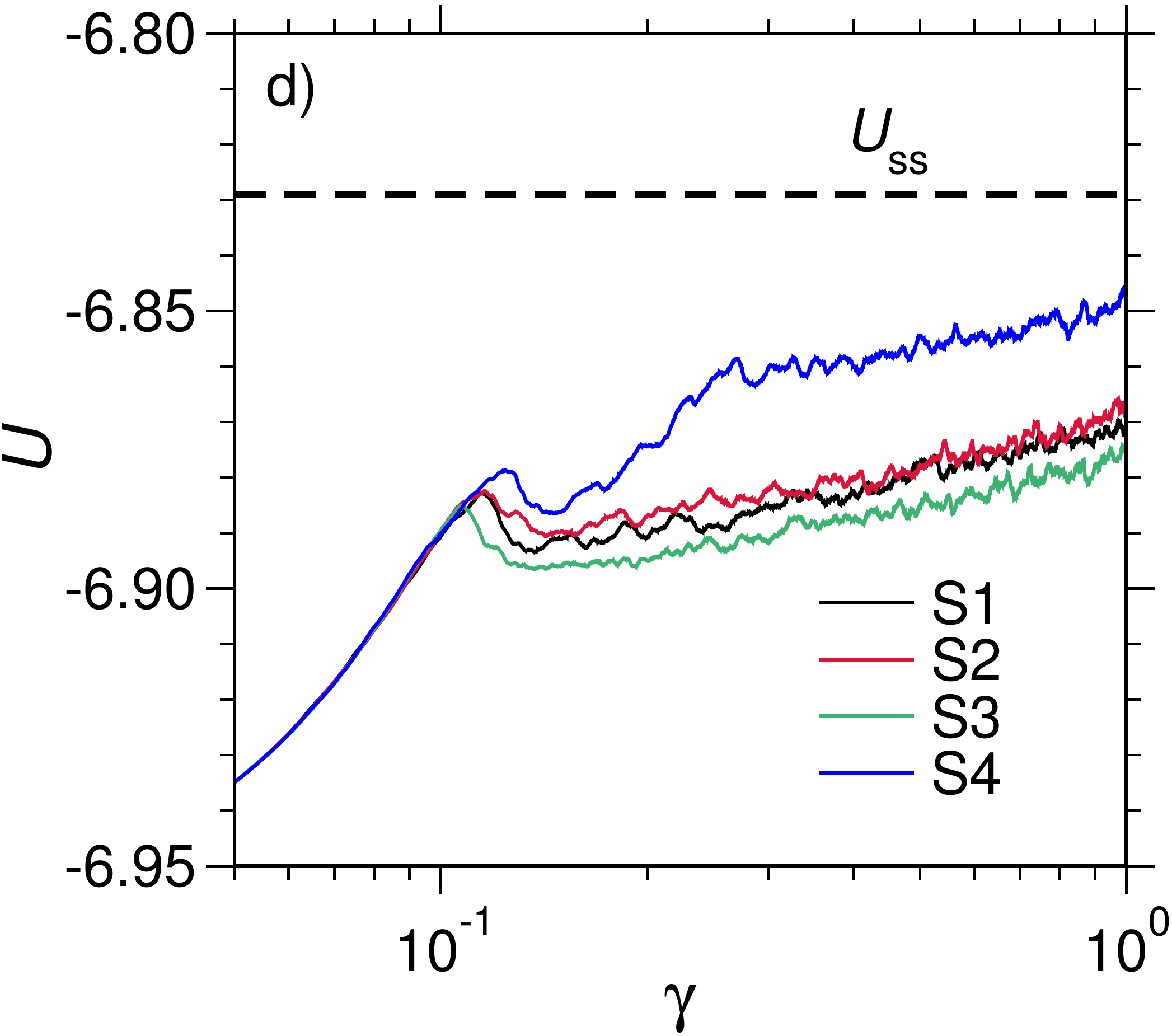}
\caption{\label{fig11} a) Snapshots of mobility color maps for four
different glass samples at $\rho = 1.3$, $T = 10^{-4}$, and
$\dot{\gamma} = 10^{-4}$ for a strain of $\gamma = 1.0$. The initial
glass configurations at $\gamma = 0$ are identical for all four
samples. b) The stress $\sigma_{xz}$, c) the $z$ component of the
squared displacement for A particles, $\delta r_{{\rm A},z}^{2}$,
and d) the potential energy, $U$, as a function of strain, $\gamma$,
for the four sheared glass samples S1 -- S4.}
\end{center}
\end{figure*}
%%%%%%%%%%%%%%%%%%%%%%%%%%%%%%%%%%%%%%%%%%%%%%%%%%%%%%%%%%%%%%

{\bf Dependence on initial glass states.} Now, the question arises
whether the type of shear band that is observed after the stress
drop around $\gamma = 0.1$ is determined by structural heterogeneities
in the initial quiescent glass sample.  To this end, we choose an
initial glass configuration and, starting from this configuration,
we perform four different simulations where we shear the system
with the same shear rate $\dot{\gamma} = 10^{-4}$, but we use in
each of these runs a different intitial random seed for the DPD
thermostat.  So only the random kicks that the particles are facing
are different in the four runs. The spatial map of squared displacements
at a strain $\gamma = 1.0$ is shown in Fig.~\ref{fig11}.  While the
first three samples have horizontal shear bands, the fourth sample
forms vertical shear bands (thus, in the latter case, the flow map
indicates an almost homogeneously flowing fluid at a strain of
$\gamma = 1.0$). Furthermore, in the first three samples, the
location of the horizontal shear band is different indicating the
stochasticity in shear band formation.

The stress-strain relations, the $z$ component of the squared
displacement for A particles, and the potential energy for the four
samples S1, S2, S3, and S4 are shown in Figs.~\ref{fig11}a,
\ref{fig11}b, and \ref{fig11}c, respectively.  With respect to these
quantities, the sample with the vertical shear band (S4) as well
as the other three with a horizontal shear band display a similar
behavior as found above for the corresponding types of shear bands.

Our findings are consistent with those of Gendelman {\it et
al.}~\cite{gendelman2015} who addressed the question whether
elementary plastic events, i.e.~shear transformation zones (STZs),
can be predicted in terms of heterogeneities in the initial unsheared
glass sample or are they characterized as events during the mechanical
load that depend on the loading protocol. To this end, Gendelman
{\it et al.}~\cite{gendelman2015} considered a glass-forming
Lennard-Jones mixture in a confined geometry with a circular shape
and found that the location of the first plastic event (STZ) depends
strongly on the details of the loading protocol. Thus, the location
of STZs cannot be simply predicted from the heterogenities of the
initial sample before the application of the mechanical load. Our
results even suggest a stochastic nature of the occurrence of plastic
events and the resulting inhomogeneous flow patterns.

%%%%%%%%%%%%%%%%%%%%%%%%%%%%%%%%%%%%%%%%%%%%%%%%%%%%%%%%%%%%%%
\begin{figure}[!htbp]
\begin{center}
\includegraphics[width=6.cm]{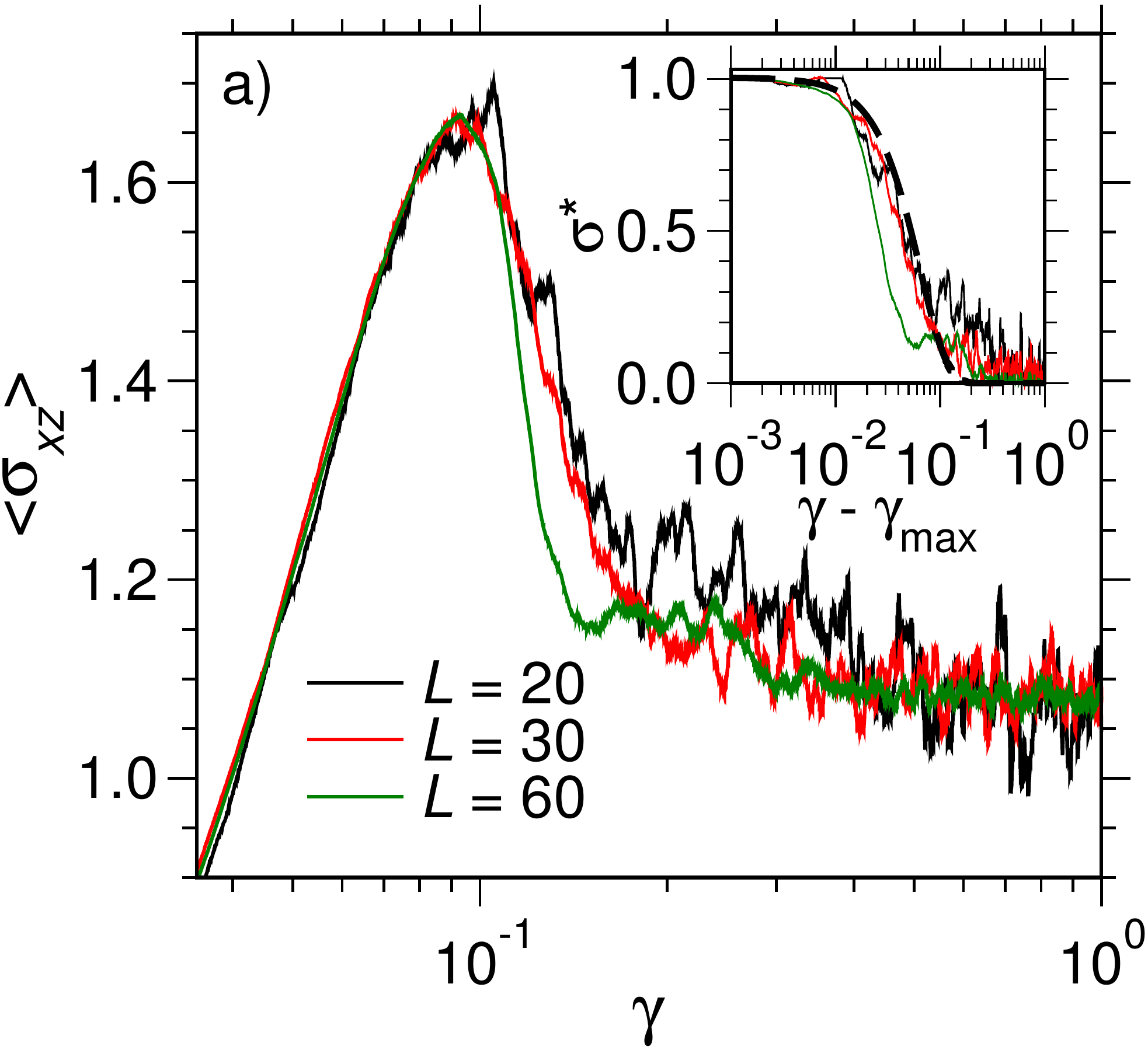}
\includegraphics[width=6.2cm]{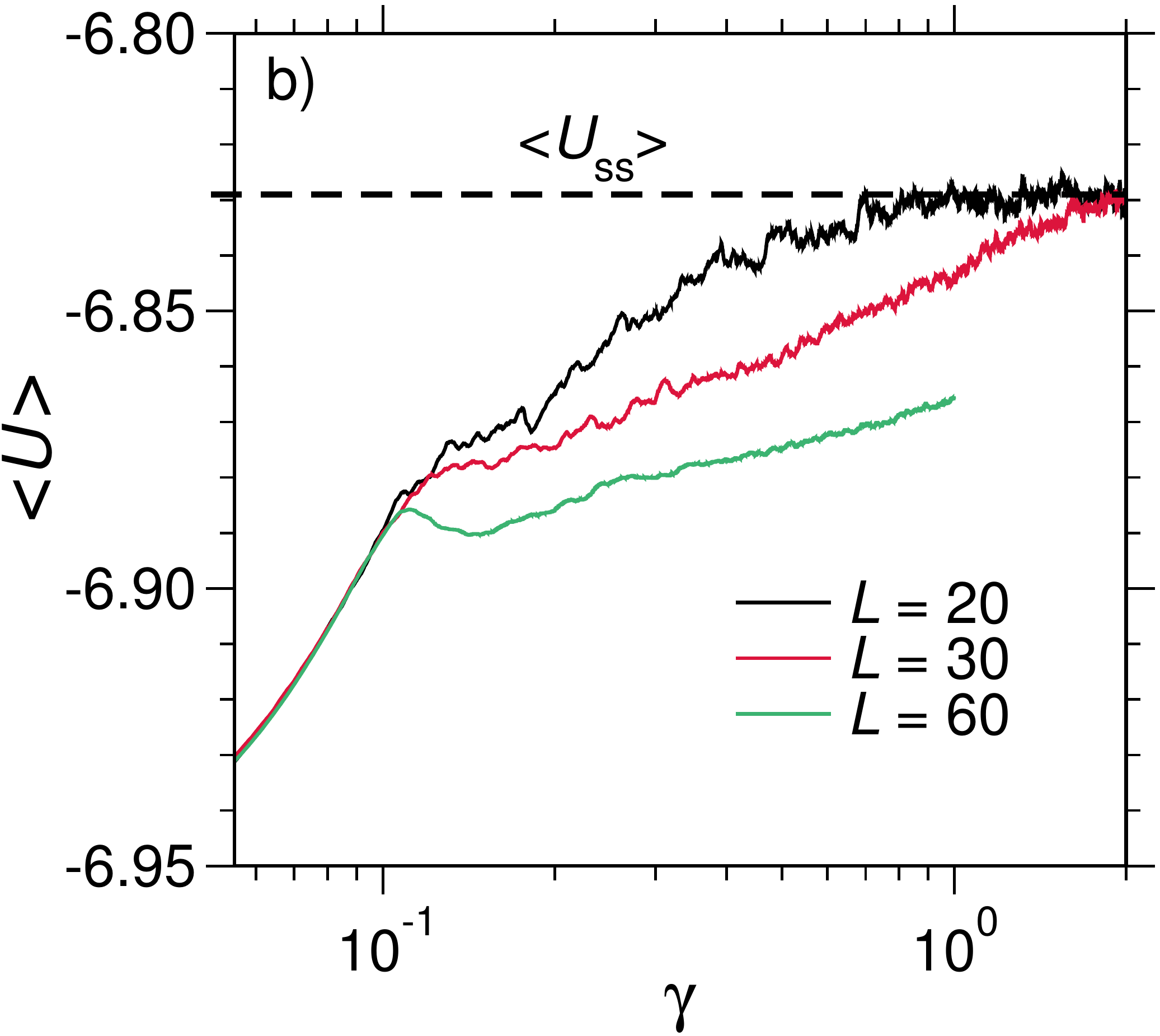}
\caption{\label{fig12} a) Stress-strain relation for three different
cubic systems with linear size $L = 20$, 30, and 60 in the main
panel and the corresponding reduced stress $\sigma^\star$ as a
function of $\gamma - \gamma_{\rm max}$ in the inset. The inset
shows in addition the compressed exponential function from
Fig.~\ref{fig1}b (dashed black line).  b) Potential energy $U$ as
a function of strain for the three different systems with $L = 20$,
30, and 60. The steady-state value $\langle U_{\rm ss} \rangle =
-6.829$ is marked by a horizontal dashed line.}
\end{center}
\end{figure}
%%%%%%%%%%%%%%%%%%%%%%%%%%%%%%%%%%%%%%%%%%%%%%%%%%%%%

{\bf Finite-size effects.} Up to now, we have shown results for
cubic glass samples with a linear size $L=60$. While in the case
of the supercooled liquids under shear we have not observed significant
finite-size effects, this is different for the low-temperature glass
states. To study finite-size effects, we analyze average properties
for systems with linear sizes $L=20$ and $L=30$ in addition to those
with $L=60$.  The averages were taken over 100, 50, and 10 samples
for $L=20$, $L=30$, and $L=60$, respectively. Figure \ref{fig12}a
shows the stress-strain relation as well as the decay of $\sigma^\star$
as a function of $\gamma - \gamma_{\rm max}$ for the different
system sizes.  These data suggest that the initial stress drop from
$\sigma_{\rm max}$ to $\sigma_{\rm ss}$ is faster for the largest
system than for the two smaller systems. This is due to a higher
probability for the emergence of horizontal shear bands in large
systems. This can be more clearly seen in the strain dependence  of
the average potential energy, $\langle U \rangle$ (Fig.~\ref{fig12}b).
The average potential energy for $L=60$ displays the behavior which
is associated with the occurrence of horizontal shear bands for
$\gamma>0.1$: a drop of $\langle U \rangle$ at $\gamma = 0.1$,
followed by a shallow minimum and a slow increase towards the
steady-state value $\langle U_{\rm ss} \rangle$. For the smaller
systems, however, $\langle U \rangle$ essentually exhibits a monotonic
and relatively fast increase towards $\langle U_{\rm ss} \rangle$.
This is due to the fact that the emergence of horizontal shear bands
is less likely for the two smaller systems. We have observed
horizontal shear bands in twelve of one hundred samples with linear
size $L=20$, in thirteen of fifty samples with $L=30$, and in four
of ten samples with $L=60$. Moreover, even when a horizontal shear
band forms in the case of small system sizes, the steady state is
reached much earlier and at comparable strains the shear banded
region covers a larger fraction of the system than for large systems.
Thus, at a given strain, the potential energy tends to increase
with decreasing system size. However, we have not found significant
finite-size effects for the average value of the potential energy
in the steady state, $\langle U_{\rm ss} \rangle$.

%%%%%%%%%%%%%%%%%%%%%%%%%%%%%%%%%%%%%%%%%%%%
\begin{figure}%[!htbp]
\begin{center}
\includegraphics[width=6.2cm]{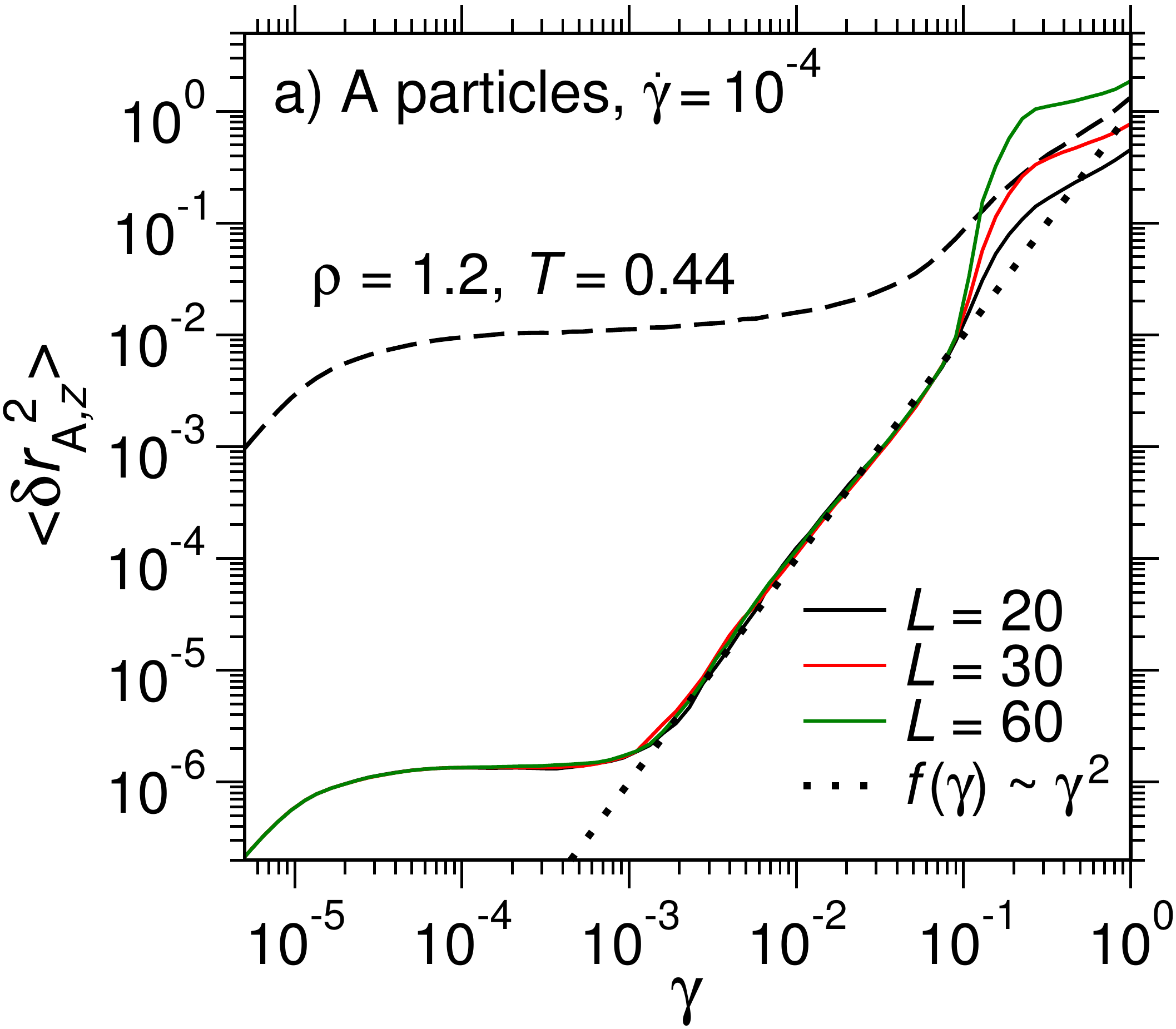}
\includegraphics[width=6.cm]{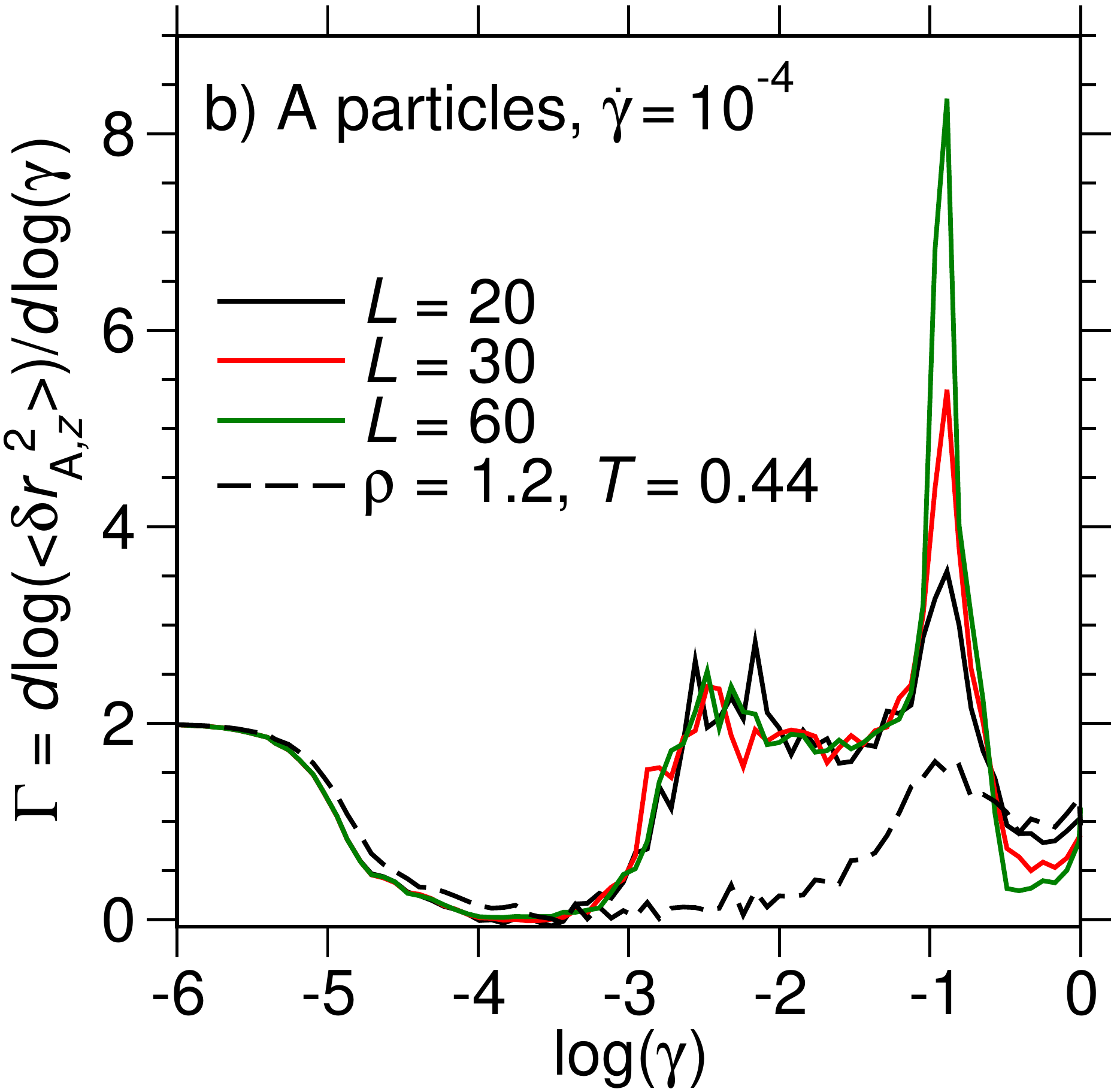}
\caption{\label{fig13} a) The $z$-component of the MSD of A particles,
$\langle \delta r_{{\rm A}, z}^2 \rangle$, as a function of strain
for the samples with $L=20$, 30, and 60 at $\rho=1.3$, $T=10^{-4}$,
and $\dot{\gamma} = 10^{-4}$. Also included is the MSD for the
supercooled liquid at $\rho = 1.2$ and $T = 0.44$. The dotted line
represents the function $f(\gamma)=a^2\gamma^2$ (with $a=1.0$).  b)
The logarithmic derivative, $\Gamma$, corresponding to the MSDs in
a).}
\end{center}
\end{figure}
%%%%%%%%%%%%%%%%%%%%%%%%%%%%%%%%%%%%%%%%%%%%%%%

We have seen that, with increasing the size of the systems with
cubic geometry, on average the stress drop in the stress-strain
relation becomes sharper. This is also reflected in the behavior
of the $z$-component of the MSD of A particles (Fig.~\ref{fig13}a).
After the ballistic regime $\langle \delta r_{{\rm A}, z}^2 \rangle
\propto \gamma^2$ in the strain range $10^{-3} < \gamma < 10^{-1}$,
the MSDs exhibit a jump that becomes more pronounced with increasing
system size.  This can be also nicely inferred from the exponent
parameter $\Gamma$, that is obtained via the logarithmic derivative
of the MSD,
\begin{equation}
\Gamma = 
\frac{d {\rm log}\left( \langle \delta r_{{\rm A}, z}^2 \rangle\right)}
{d {\rm log} \langle(\gamma \rangle)} \, .
\end{equation}
This quantity is shown Fig.~\ref{fig13}b for the different system
sizes.  Here, we can identify the short-time ballistic regime with
$\Gamma=2$ for $\gamma < 10^{-6}$, the plateau region with $\gamma
\approx 0$ for $10^{-4}< \gamma < 10^{-3}$, and the ballistic regime
$\propto \gamma^2$ with $\Gamma \approx 2$ for $10^{-3}< \gamma <
10^{-1}$. In the latter regimes, no finite-size effects are observed.
However, for $\gamma \approx 0.1$, there is a peak in $\Gamma({\rm
log} \gamma)$ that increases with system size up to a value of
$\Gamma \approx 8.0$ for the system with $L=60$.  So the stress
drop in the stress-strain relation is linked to a jump in the MSD
which becomes more pronounced with increasing system size.  We note
that a similar behavior, albeit much more pronounced, has been
recently found by Ozawa {\it et al.}~\cite{ozawa2018}, using an
aqs protocol.

\section{Summary and conclusions}
\label{sec_conclusions}
We have performed non-equilibrium MD simulations of a glassforming
binary Lennard-Jones mixture under shear. The shear response
of supercooled liquid states has been compared to that of glass
states at extremely low, albeit finite, temperature, with the focus
on the characterization of yielding and inhomogeneous flow patterns
(especially shear bands).

In the supercooled liquid state, we have identified a critical shear
rate $\dot{\gamma}_c$ that marks the crossover from Newtonian to
non-Newtonian response of the liquid to the external shear. For
$\dot{\gamma} > \dot{\gamma_c}$, i.e.~in the non-Newtonian regime,
we find inhomogeneous flow patterns in the supercooled liquid in
the form of vertical bands.  These bands are short-lived and are
observed right after the overshoot in the stress-strain relation.
The decay of the stress from the maximum stress at the overshoot
to the steady-state stress $\sigma_{\rm ss}$ does not seem to depend
on shear rate at a given temperature (at least over a large range
of shear rates). No finite-size effects are seen in the stress-strain
relation, too. The characteristic strain window over which the
stress is released is of the order of $\Delta \gamma = 0.1$.  The
potential energy shows a step-like behavior when the system yields,
i.e.~a monotonic increase again over a strain window of about
$\Delta \gamma = 0.1$ towards the steady-state value.

In the glass, we have studied the response of glass samples having
similar annealing history at a given shear rate, $\dot{\gamma} =
10^{-4}$, and temperature, $T=10^{-4}$.  For these samples, we
observe inhomogeneous flow patterns that differ from sample to
sample.  In some of them, the plastic flow is associated with
relatively short-lived vertical shear bands (i.e.~with an orientation
perpendicular to the flow direction), while in other samples
horizontal bands are seen (i.e.~aligned with the flow direction).
Also mixed patterns with vertical and horizontal bands are observed.
The type of flow pattern that is seen in the transient plastic flow
regime is not predetermined by the structure of the initial un-sheared
sample.  Starting from the same configuration but with different
random numbers for the velocity distribution may lead to vertical
or to horizontal flow patterns. This indicates the stochasticity
of the process. For the sheared glass samples, the behavior of the
potential energy is different from that of the supercooled liquids
in the non-Newtonian regime.  Now, this quantity is not always
monotonically increasing towards the steady-state stress.  For the
cases, where horizontal shear bands occur, the potential energy is
higher in the shear band than in the other regions in the system.
In the latter regions, it drops to values that are significantly
below the final steady-state value.  So the system finds a state
of lower energy, however this state is not stable in the presence
of the applied deformation and so the horizontal shear band broadens
as a function of time in a subdiffusive manner. Moreover, in the
case of horizontal bands, the stress drop in the stress-strain
relation is sharper than that in the supercooled state and it becomes
sharper with increasing system size. Furthermore, via a spatial
resolution of the system's potential energy, we show that it is
possible to identify the strain value at which horizontal shear-bands
emerge, and this occurs beyond the strain where the stress overshoot
is observed.

We have seen that there are similarities between the shear response
of a supercooled liquid in the non-Newtonian regime and a glass.
In both cases, the generic behavior for sufficiently low shear rates
is as follows: There is first a strong elastic response to the shear
for strains $\gamma<0.1$. This results in a deformed amorphous solid
(note that for $\gamma<0.1$ also the supercooled non-Newtonian
liquid exhibits solid-like behavior). Then, after a stress release
as reflected by a stress drop in the stress-strain relation, the
deformed solid eventually transforms into a homogeneously flowing
state that can be characterized as an anisotropic non-equilibrium
fluid. The latter fluid state is a well-defined stationary state
that is obtained at a given shear rate, temperature, and density
of the system; in particular, it is independent of the history of
the initial unloaded state from which it was obtained via a certain
shear protocol. For the sheared glass, the pathways with which the
stationary fluid state is reached can completely differ from sample
to sample. In the cases with horizontal shear bands one observes a
kind of nucleation of the flowing fluid phase that grows slowly
towards the homogeneously flowing stationary state.

So the yield point, when identified with the maximum in the
stress-strain relation, marks the onset of plastic flow, but it
describes a ``transition'' to a transient state that evolves into
a homogeneous fluid state. From this point of view, it is interesting
that inhomogeneous states with horizontal shear bands can be
stabilized via oscillatory shear. This has been recently shown in
two independent simulation studies~\cite{parmar2019, miyazaki2019}.
These works have found stationary horizontal shear bands when the
maximum strain amplitude in the oscillatory shear is above the
critical yield strain, corresponding to the maximum in the stress-strain
relation.  As a result, an inhomogeneous system is obtained where
a fluidized shear-banded region coexists with a glassy region. With
respect to potential energy, these states look similar to the ones
with horizontal shear bands that we find at a fixed strain
(cf.~Fig.~\ref{fig9}b). However, it is not clear whether the
inhomogeneous state that one observes in the oscillatory shear
simulation corresponds to a ``true'' stationary state. If one were
able to perform more cycles, the glassy region outside the shear
band might further age, as indicated by subdiffusive relaxation
processes in Ref.~\cite{parmar2019}.

While one fixes the window of allowed strains in simulations using
an oscillatory shear protocol, one can also perform simulations of
glasses subjected to a constant external stress \cite{chaudhuri2013,
cabriolu2019}.  As in similar experiments \cite{siebenbuerger2012},
one obtains the strain $\gamma$ as a function of time $t$ from such
simulations.  If the external stress is below the yield stress,
there is only creep flow and the strain shows a sublinear increase
of the strain as a function of time \cite{chaudhuri2013, cabriolu2019,
siebenbuerger2012}.  In Ref.~\cite{chaudhuri2013}, it was shown
that this creep flow is associated with the formation of a shear-banded
region. Again, this inhomogeneous system with a shear band does not
correspond to a stationary state. This is due to the fact that the
time scales required to reach a steady state with $\gamma \propto
t$ are not accessible by the simulation.

An important issue is the role of long-range strain correlations
\cite{barrat2011} for the formation of horizontal shear bands. Such
correlations are a characteristic feature of elastic media and are
also present in supercooled liquids \cite{chattoraj2013, hassani2018}.
The mechanisms of how strain correlations are connected to the
formation of inhomogeneous flow patterns is not well understood.
It is a subject of forthcoming studies to elucidate this issue.

\begin{acknowledgments}
We thank Kirsten Martens, Vishwas Vasisht, Misaki Ozawa, and Surajit Sengupta for useful discussions.  The authors
acknowledge the financial support by the Deutsche Forschungsgemeinschaft
(DFG) in the framework of the priority programme SPP 1594 (Grant
No.~HO 2231/8-2). One of us (G.~P.~S.) acknowledges financial support
by the Austrian Science Foundation (FWF) under Proj.~No.~I3846.
\end{acknowledgments}


\begin{thebibliography}{99}
\bibitem{fn1}
That we give the fit parameter with seven decimal places is just for the
purpose of reproducibility of the fit function.
\end{thebibliography}
\end{document}